\documentclass[aps,prb,twocolumn,groupedaddress,amsmath,amssymb,showpacs]{revtex4-1}
 \usepackage{graphicx,graphics,color}
\usepackage{amsmath}
\usepackage{amssymb}
\usepackage{amsfonts}
\usepackage{amsfonts}
\usepackage{epstopdf}
\usepackage{bm}
\usepackage{times,xspace}
\usepackage{color}
\usepackage[utf8]{inputenc}

\def\be{\begin{equation}}
\def\ee{\end{equation}}
\def\bea{\begin{eqnarray}}
\def\eea{\end{eqnarray}}

\def\PT{\mathcal{PT}}

\def\T{\mathcal{T}}

\def\CPT{\mathcal{CPT}}

\def\PT{$\mathcal{PT}$}
\def\T{$\mathcal{T}$}

\begin{document}

\title{Theory of superconductivity with non-hermitian and parity-time reversal symmetric cooper pairing symmetry}

\author{Ananya Ghatak\footnote{gananya04@gmail.com}, and Tanmoy Das\footnote{tnmydas@gmail.com}\\
{Department of Physics, Indian Institute of Science, Bangalore-560012, India.}}

\date{\today}
\vspace{0.3cm}

\begin{abstract}
Recently developed parity ($\mathcal{P}$) and time-reversal ($\mathcal{T}$) symmetric non-Hermitian systems govern a rich variety of new and characteristically distinct physical properties, which may or may not have a direct analog in their Hermitian counterparts. We study here a non-Hermitian, $\mathcal{PT}$-symmetric superconducting Hamiltonian that possesses real quasiparticle spectrum in the $\mathcal{PT}$-region of the Brillouin zone. Within a single-band mean-field theory, we find that real quasiparticle energies are possible when the superconducting order parameter itself is either Hermitian or {\it anti-Hermitian}. Within the corresponding Bardeen-Cooper-Schrieffer (BCS) theory, we find that several properties are characteristically distinct and novel in the non-Hermitian pairing case than its Hermitian counterpart. One of our significant findings is that while a Hermitian superconductor gives a second order phase transition, the non-Hermitian one  produces a robust first order phase transition. The corresponding thermodynamic properties, and the Meissner effect are also modified accordingly. Finally, we discuss how such a $\mathcal{PT}$-symmetric pairings can emerge from an {\it anti-symmetric} potential, such as the Dzyloshinskii-Moria interaction, but with an external external bath, or complex potential, among others.
\end{abstract}
\maketitle

\section{Introduction}\label{intro}

The $\mathcal{PT}$-symmetric (where $\mathcal{P}$ and $\mathcal{T}$ are parity and time reversal operators, respectively) class of non-Hermitian (NH) systems with real energy eigenvalues have become the topic of frontier research over a decade and half \cite{ben4,benr,mos,PToperator1,PToperator2}. The Hermiticity requirement of a Hamiltonian is replaced by the analogous condition of $\mathcal{PT}$ symmetry, and with this, one can have a consistent quantum theory with a unitary time evolution\cite{ben4,benCopp}. It is anticipated that such $\mathcal{PT}$-symmetric NH systems can govern new and exotic physical properties, which may or may not have direct analogs in the Hermitian counterparts. Recent experimental realizations of such Hamiltonians in condensed matter systems such as optical systems\cite{opt1,eqv1}, and metamaterials\cite{Alaeian,PTmat} have provided a huge boost to this field, and hitherto studies of such theories have dispersed into various branches of physics \cite{ent,cal}. For example, various topological properties of the $\mathcal{PT}$-symmetric NH Hamiltonian are recently investigated \cite{NHTI1,NHTI2}. Stability and localization of various normal state and superconducting properties under NH disorder have also been studied recently \cite{Nsc1,Nsc4}.

Interestingly, it is recently observed that superconductivity is significantly enhanced in metamaterials and optically pumped cuprates\cite{SCMetamatSn, SCMetamat,SCMetamatAl, OpticalSC}, where also non-Hermiticity may concur \cite{opt1,eqv1,Alaeian,PTmat}. While we draw physical motivation to study a NH Cooper pairing instability from these observations, but the corresponding theory is generic and is also applicable to other systems. $\mathcal{P}$ and/or $\mathcal{T}$ broken superconductors have been discussed in non-centrosymmetric materials,\cite{non-cento1,non-cento2,non-cento3} and in odd-frequency pairing cases,\cite{odd-freq1,odd-freq2,odd-freq3} but to our knowledge, the combined $\mathcal{PT}$-invariant pairing symmetry has not been studied before even for a Hermitian case.

We start with delineating the general properties of a NH superconducting state which can describe a physical system. Superconductivity arises when the low-energy electrons and holes individually pair up, owing to an effective attractive potential between them. The superconducting (SC) gap due to the electron-electron and hole-hole pairs, respectively, are $\tilde{\Delta}_{\bf k}=-\sum_{{\bf k}'}V_{{\bf k}{\bf k}'}\langle c^{\dag}_{{\bf k}'\uparrow}c^{\dag}_{-{\bf k}'\downarrow}\rangle$, and $\Delta_{\bf k}=-\sum_{{\bf k}'}V_{{\bf k}'{\bf k}}\langle c_{-{\bf k}'\downarrow}c_{{\bf k}'\uparrow}\rangle$, where $V_{{\bf k}{\bf k}'}$ is the pairing potential, $c^{\dag}_{{\bf k}\uparrow}$, and $c_{{\bf k}\uparrow}$ are the creation and annihilation operators of electrons at momentum ${\bf k}$ with spin up. In Hermitian superconductors, $\tilde{\Delta}_{\bf k}=\Delta_{\bf k}^{\dag}$. Here investigate a generic case where such a constrain is relaxed, and replaced with a more generalized criterion that these two pairs are \PT conjugate to each other, i.e., $\tilde{\Delta}_{\bf k}=\Delta_{\bf k}^{\mathcal{PT}}$, but not necessarily Hermitian. For a non-interacting dispersion $\varepsilon_{\bf k}$ (real function), the BCS energy eigenvalues are $E_{\bf k}=\pm \sqrt{\varepsilon_{\bf k}^2+\tilde{\Delta}_{\bf k}\Delta_{\bf k}}$. It is convenient to express the complex gaps in polar coordinates as $\Delta_{\bf k}=|\Delta_{\bf k}|e^{i\theta_{\bf k}}$, and $\tilde{\Delta}_{\bf k}=|\tilde{\Delta}_{\bf k}|e^{i\tilde{\theta_{\bf k}}}$, where the modulus of the gap (real) is defined as $|\Delta_{\bf k}|^2=\Delta_{\bf k}\Delta_{\bf k}^{\dag}$, and $\theta_{\bf k}$, $\tilde{\theta}_{\bf k}$ are the corresponding phases. Energies are real when $\tilde{\Delta}_{\bf k}\Delta_{\bf k}=|\tilde{\Delta}_{\bf k}||\Delta_{\bf k}|e^{i(\tilde{\theta}_{\bf k}+\theta_{\bf k})}$ is real. This is achieved when the two phases follow $\tilde{\theta}_{\bf k}+\theta_{\bf k}=n\pi$, with $n\in \mathbb{Z}$ ($n$ can be ${\bf k}$-dependent, but we take isotropic case for simplification). There arises two different scenarios when $n$ assumes either even or odd integers. 

When $n$ is {\it even}, we obtain $\tilde{\Delta}_{\bf k}\Delta_{\bf k}=|\tilde{\Delta}_{\bf k}||\Delta_{\bf k}|$. $|\tilde{\Delta}_{\bf k}|=|\Delta_{\bf k}|$ condition gives a Hermitian system, whereas $|\tilde{\Delta}_{\bf k}|\ne|\Delta_{\bf k}|$ produces a NH superconductor. In addition, the order parameter is assumed to be \PT-symmetric, so we have $|\Delta_{\bf k}|=|\Delta_{\bf k}^{\mathcal{PT}}|=|\tilde{\Delta}_{\bf k}|$ (since modulus gives a real number which is invariant here). In what follows, one {\it cannot} obtain a \PT-symmetric, NH SC Hamiltonian with precisely opposite phases (i.e., when $n$ is even). 

On the contrary, when $n$ is {\it odd}, we obtain $\tilde{\Delta}_{\bf k}\Delta_{\bf k}=-|\tilde{\Delta}_{\bf k}||\Delta_{\bf k}|$. This suggests that, if we take $\Delta_{\bf k}=|\Delta_{\bf k}|e^{i\theta_{\bf k}}$, then $\tilde{\Delta}_{\bf k}=-|\tilde{\Delta}_{\bf k}|e^{-i\theta_{\bf k}}$. Employing the the \PT-invariance condition (i.e., $|\tilde{\Delta}_{\bf k}|=|\Delta_{\bf k}|$), we get $\tilde{\Delta}_{\bf k}=-\Delta_{\bf k}^{\dag}$, which makes the SC gap {\it anti-Hermitian} (but owing to the $\xi_{\bf k}$ term, the Hamiltonian is not anti-Hermitian, but NH). In other words, the \PT-invariance implies that $\mathcal{PT}e^{i\theta_{\bf k}}(\mathcal{PT})^{-1} = -e^{-i\theta_{\bf k}}$. Hence $\theta_{\bf k}\ne 0$, which excludes the possibility of a purely real order parameter in the \PT-symmetric, NH-SC case. 

Based on the aforementioned properties, we can construct the BCS theory for such a generic NH Hamiltonian with $\mathcal{PT}$-symmetric pairings. We compare the results with those of a corresponding Hermitian superconductor with the same \PT-symmetric pairing symmetry. These two cases, as refereed to `NH-SC' and `H-SC' Hamiltonians, are defined as
\bea
\textrm{H-SC}:&&~~~\tilde{\Delta}_{\bf k}=\Delta_{\bf k}^{\mathcal{PT}}~~\&~~\tilde{\Delta}_{\bf k}=\Delta_{\bf k}^{\dag},\nonumber\\
\textrm{NH-SC}:&&~~~\tilde{\Delta}_{\bf k}=\Delta_{\bf k}^{\mathcal{PT}}~~\&~~\tilde{\Delta}_{\bf k}=-\Delta_{\bf k}^{\dag}.
\label{DefineNH}
\eea
We summarize our general result in Table~\ref{tableA}. For the H-SC case, the eigenvalues $E_{\bf k}=\pm \sqrt{\varepsilon_{\bf k}^2+|\Delta_{\bf k}|^2}$ are real at all ${\bf k}$ points on the Brillouin zone (BZ). On the other hand, the quasiparticle energy for the NH-SC $E_{\bf k}=\pm \sqrt{\varepsilon_{\bf k}^2-|\Delta_{\bf k}|^2}$ is real in the \PT-invariant (`paired') region where $|\varepsilon_{\bf k}|\ge|\Delta_{\bf k}|$, while in outside, the quasiparticle states break the \PT- symmtry and thus the superconductivity remains blocked (namely `unpaired' region). The Free-energy in the leading of the gaps takes the form $F_s-F_n=a\tilde{\Delta}\Delta=\pm a|\Delta|^2$ for the H-SC and NH-SC cases, respectively ($F_n$ includes the non-SC contributions). For the H-SC case, it becomes minimum when $a<0$, giving a typical second order phase transition. On the other hand, for the NH-SC case, the Free energy is lowered for $a>0$, which we will show below, within the Ginsburg-Landau theory, that it gives a {\it first} order phase transition. 

\begin{table}[t]
\begin{tabular}{|c |c |c|}
\hline
{\bf Properties} & {\bf H-SC} & {\bf NH-SC} \\
\hline
\hline
SC gap &  $\tilde{\Delta}_{\bf k}=\Delta_{\bf k}^{\dag}$, & $\tilde{\Delta}_{\bf k}=-\Delta_{\bf k}^{\dag}$ \\
\hline
Eigenvalues &  $\pm \sqrt{\varepsilon_{\bf k}^2+|\Delta_{\bf k}|^2}$ & $\pm \sqrt{\varepsilon_{\bf k}^2-|\Delta_{\bf k}|^2}$ \\
\hline
Free energy & $F_s-F_n=a|\Delta|^2$ & $F_s-F_n=-a|\Delta|^2$ \\
\hline
Phase transition & Second order & First order\\
\hline
Pairing interaction & Symmetric & Anti-symmetric \\
                          & ($V_{\bf kk'}=V_{\bf k'k}$) & ($V_{\bf kk'}=-V_{\bf k'k}$) \\
\hline
\end{tabular}
\caption{The table gives a comparison between the two cases where energy eigenvalues are real even without a Hermitian operator, but with \PT invariance. $F_n$ is the normal state Free-energy.
 }
\label{tableA}
\end{table}

In addition, we also show that the NH order parameter can, for example, emerge from an anti-symmetric potential $V_{\bf kk'}=-V_{\bf k'k}$ (or anti-Hermitian, if complex). Dzyaloshinskii-Moriya (DM) interaction which arises in non-centrosymmetric systems is one such an anti-symmetric, real potential. It can give an anti-Hermitian pair, if the system is connected to a bath or the potential is made complex. This suggests that a NH-SC order parameter can emerge in a physical system even from a Hermitian normal state. Finally, we find that the self-consistent gap function, thermodynamical, and transport properties turned out to be characteristically different here compared to the Hermitian case with the same pairing function. We reaffirm the characteristic differences between the type of phase transitions in both H-SC and NH-SC cases with self-consistent gap calculation within the BCS theory. 

The rest of the manuscript is arranged as follows. In Sec.~\ref{Sec:Model}, we describe the one-band model with a \PT-symmetric Hermitian and anti-Hermitian SC order parameters, their differences in eigenstates, and ground state properties. We also discuss the definition of the $\mathcal{CPT}$ inner products and expectation values of physical properties in the NH-SC state. In Sec.~\ref{Sec:Results}, we present the self-consistent gap equation, Free energy calculations, the Ginsburg-Landau description of the phase transition, and the Meissner effect. We compare all the results for both the Hermitian and non-Hermitian cases with the same pairing symmetry. Finally, we discuss various aspects of the model, results, and the possibility of their realizations in condensed matter systems in Sec.~\ref{Sec:Discussion}. In Appendix~\ref{DM}, we derive the non-Hermitian SC order from a Hermitian DM interaction.

\section{Model}\label{Sec:Model}

\subsection {$\mathcal{PT}$-symmetric order parameter}\label{Sec.IIA}

The theory of $\mathcal{PT}$ symmetric Hamiltonian suggests that all complex conjugate terms are replaced by their corresponding $\mathcal{PT}$-conjugate, i.e. $c^{\dag}_{{\bf k}\sigma}\rightarrow c^{\mathcal{PT}}_{{\bf k}\sigma}$. For generalization, we henceforth use the symbol `tilde' to denote `dagger' for a Hermitian case, and $\mathcal{PT}$-conjugate for the NH counterpart. Using this convention, we start with a generalized pairing Hamiltonian as,
\be
H=\sum_{{\bf k}\sigma } \varepsilon _{\bf k} \tilde{c}_{{\bf k}\sigma}c_{{\bf k}\sigma }+\sum_{{\bf k,k'}} V_{{\bf kk'}} \tilde{c}_{{\bf k}\sigma }\tilde{c}_{-{\bf k}\bar{\sigma}}  c_{-{\bf k'}\bar {\sigma} }c_{{\bf k'}\sigma },
\label{HSC1}
\ee
where $\tilde{c}_{{\bf k}\sigma}$ ($c_{{\bf k}\sigma}$) is the creation (annihilation) operator for an electron with Bloch momentum ${\bf k}$, and spin $\sigma$, with $\bar{\sigma}=-\sigma$ for singlet- and $\bar{\sigma}=\sigma$ for triplet pairings. The non-interacting dispersion $\varepsilon_{\bf k}$ is considered within the tight-binding model with nearest neighbor hopping ($t$) as $\varepsilon _{{\bf k}}=-2t(\cos{k_x}+\cos{k_y})-\mu$, and $\mu$ is the chemical potential. We set $t=1$, and the Boltzmann constant $k_B=1$ in all the calculations and figures below. The pairing potential $V_{\bf kk'}$ is general, i.e., it can arise from either electron-phonon coupling, or electronic interaction and the discussion of the potential is deferred to Appendix~\ref{DM}.

We define the pair creation and annihilation operators (the SC fields) as 
\bea
\tilde{\phi}_{{\bf k}} =\tilde{c}_{{\bf k}\sigma }\tilde{c}_{-{\bf k}\bar{\sigma}}, ~~~\phi_{{\bf k}} =c_{-{\bf k}\bar{\sigma}}c_{{\bf k}\sigma}.
\label{phi}
\eea
Let us assume $\langle \phi_{{\bf k}} \rangle_{\CPT}$, and $\langle \tilde{\phi}_{{\bf k}} \rangle_{\CPT}$ are the two corresponding mean-field values obtained from $\mathcal{CPT}$-expectation values (see Sec.~\ref{Sec.R1}). Since, all inner products are understood to be  a $\CPT$ inner product for the NH case and a typical inner product for the H-case, and we drop the superscript henceforth for simplicity. Now expanding the fields with respect to their corresponding mean-values as $\phi_{{\bf k}\sigma}=\langle \phi_{{\bf k}\sigma} \rangle + \delta \phi_{{\bf k}\sigma}$, we obtain the pairing interaction term from Eq.~\eqref{HSC1}, 
\bea
H_{\rm I}&\approx&\sum_{{\bf k}{\bf k}'\sigma} V_{{\bf k}{\bf k}'}\big(\langle \tilde{\phi}_{{\bf k}\sigma}\rangle\delta \phi_{{\bf k}'\bar{\sigma}}
+ \langle \phi_{{\bf k}'\bar{\sigma}} \rangle \delta \tilde{\phi}_{{\bf k}\sigma} \nonumber\\
&&\quad\qquad + \langle \tilde{\phi}_{{\bf k}\sigma}\rangle\langle \phi_{{\bf k}'\bar{\sigma}} \rangle 
+ \delta \tilde{\phi}_{{\bf k}\sigma}\delta \phi_{{\bf k}'\bar{\sigma}}\big).
\label{hsc2}
\eea
The order parameters are defined as
\be
\tilde{\Delta}_{\bf k} = -\sum_{{\bf k}'}V_{\bf k'k}\langle \tilde{\phi}_{{\bf k'}}\rangle,~~~\Delta_{\bf k} = -\sum_{{\bf k}'}V_{\bf kk'}\langle {\phi}_{{\bf k'}}\rangle.
\label{SCGap}
\ee
This gives the generalized BCS Hamiltonian (neglecting the last term being small, and ignoring the third being constant), 
\be
H_{\rm MF}=\sum_{{\bf k}\sigma }\left[ \varepsilon _{\bf k} \tilde{c}_{{\bf k}\sigma }c_{{\bf k}\sigma } -\Delta_{\bf k} \tilde{c}_{{\bf k}\sigma }\tilde{c}_{-{\bf k}\bar{\sigma} } -\tilde{\Delta}_{{\bf k}} c_{-{\bf k}\bar{\sigma} }c_{{\bf k}\sigma }\right].
\label{BCSHamiltonian}
\ee
Using Eqs.~\eqref{DefineNH} and \eqref{SCGap}, we can deduce the symmetry requirement for $V_{\bf kk'}$ for the two cases.  For the H-SC case, $\tilde{\Delta}_{\bf k}= \Delta_{\bf k}^\dagger \Rightarrow V_{\bf kk'}=V_{\bf k'k}^*$. On the other hand, for the NH-SC case, we have $\tilde{\Delta}_{\bf k}=-\Delta_{\bf k}^\dagger \Rightarrow V_{\bf kk'}=-V_{\bf k'k}^*$, i.e., the pairing potential matrix must be {\it anti-Hermitian} if complex, or simply {\it anti-symmetric} if real. In Appendix~\ref{DM}, we show that the DM interaction\cite{DM} can give a \PT symmetric anti-Hermitian pairing, by relaxing either the momentum or the particle-number conservation principles, or with a complex potential (see Sec.~\ref{Sec:Discussion}(vi) for specific discussions). We note that although the pairing term is anti-Hermitian, the mean-field Hamiltonian in Eq.~\eqref{BCSHamiltonian} is not anti-Hermitian, but non-Hermitian.

All the analytical formulas derived in this work does not assume any particular form of the order parameter. However, only for numerical calculations, we need to invoke a pairing symmetry which is kept fixed for both H-SC and NH-SC cases for direct comparison. In the single band case, the anti-Hermiticity implies that the order parameter is purely imaginary, breaking the \T-symmetry. Therefore, to preserve the \PT symmetry, the order parameter must be odd under parity. So, we consider a $id_{xy}$-pairing symmetry as 
\be
\Delta_{{\bf k}}=i\Delta _0 \sin{(k_x)}\sin{(k_y)},
\label{pairingsymmetry}
\ee
where $\Delta_0$ is the SC gap amplitude. In 2D, the parity operator is defined with respect to a mirror plane as \cite{adas_2d} $\mathcal{P}: (x,y) \longrightarrow (-x,y)$, or $(x,y) \longrightarrow (x,-y)$.\cite{dimension} Then under an usual $\mathcal{T}$ operator, we find that the gap in Eq.~\eqref{pairingsymmetry} satisfies $\mathcal{P}\mathcal{T} \Delta_{{\bf k}}(\mathcal{P}\mathcal{T})^{-1}=\Delta_{{\bf k}}$. we have also studied other forms of the \PT-symmetric order parameters in Appendix~\ref{Appen-E}, and we have found that the general conclusions remain the same.

\subsection{Eigenvalues and eigenfunction}\label{Sec.IIB}

\begin{figure}[t]
\centering
\includegraphics[width=.6\columnwidth]{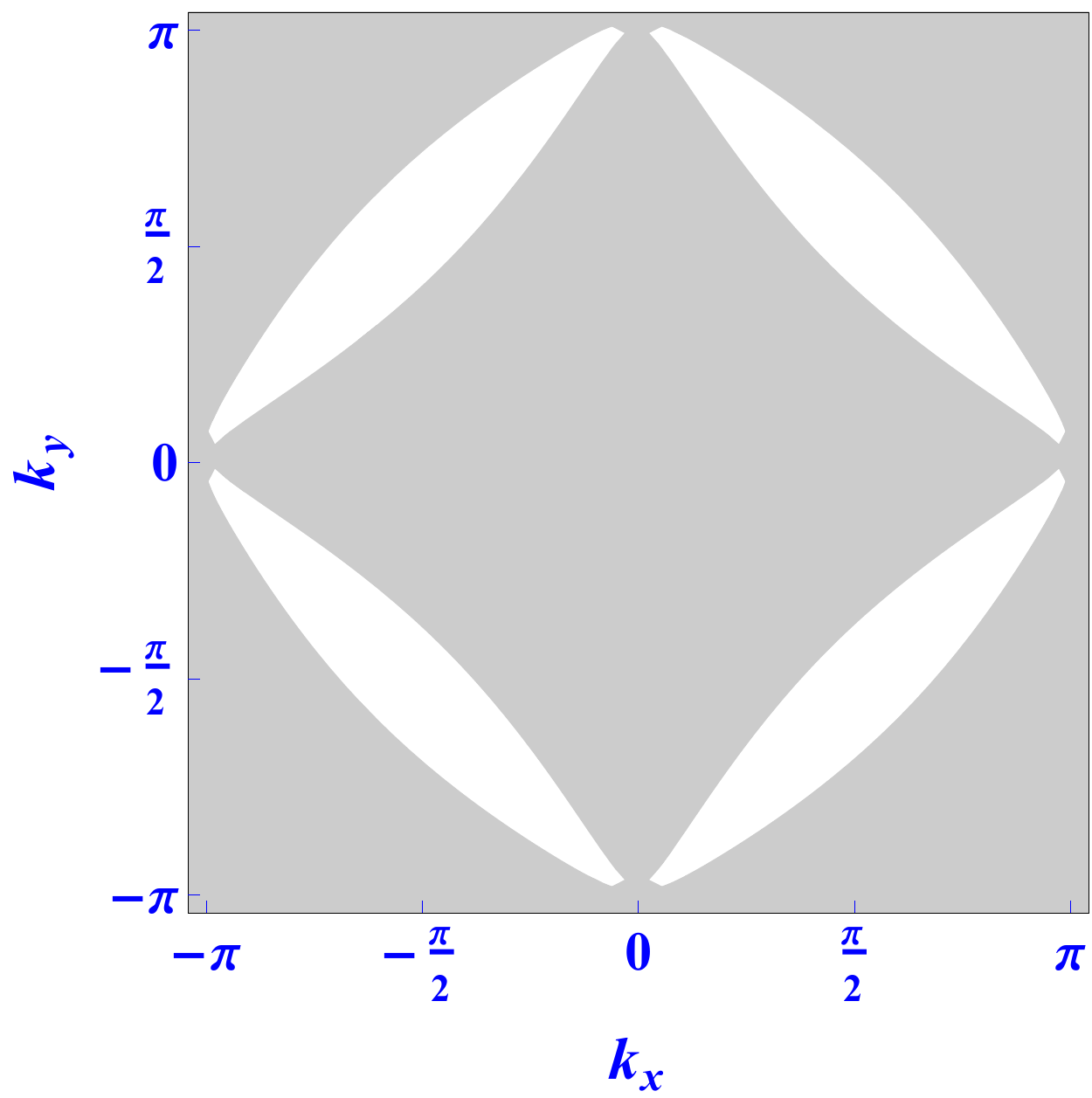}
\vspace{-.14in}
\caption{The splitting of the BZ into the `paired region' (gray color), and `unpaired region' (white) in the NH-SC case. The shape of the `unpaired region'  is determined by the pairing symmetry at hand, while the area is proportional to the gap amplitude ($\Delta_0$). As $T \rightarrow  T_c$ the white region vanishes smoothly. $k_{x,y}$ are defined in units of $1/a$ where $a$ is the lattice constant. }
\label{fig1}
\end{figure}

The eigenvalues of Eq.~\ref{BCSHamiltonian} are given by $E_{{\bf k}}=\pm\sqrt{\varepsilon _{{\bf k}}^2+\Delta_{{\bf k}}\tilde{\Delta}_{{\bf k}}}=\pm\sqrt{\varepsilon _{{\bf k}}^2+|\Delta _{\bf k}|^2}$, for the H-SC system, and $E_{{\bf k}}=\pm\sqrt{\varepsilon_{{\bf k}}^2-|\Delta _{\bf k}|^2}$ for the NH-SC case. Clearly, in the latter case, the eigenvalues are real only in the region, called `paired region' ($\mathfrak{R_1}$), defined by the boundary, 
\be
|\varepsilon _{{\bf k}}|\geq |\Delta _{\bf k}|,
\label{ptcon}
\ee 
as depicted by grey shadings in Fig.~\ref{fig1}.  The white region is called `unpaired region' ($\mathfrak{R_2}$) where \PT symmetry is broken, and the SC quasiparticle states rapidly decay to the normal state. As the SC gap $|\Delta_{\bf k}|$ decreases, the size of the `paired region' gradually increases, and it smoothly covers the entire BZ at $T_c$. Such an `unpaired region' also arises in H-SC Hamiltonians with finite-momentum pairings, as referred by Fulde–Ferrell–Larkin–Ovchinnikov (FFLO) phase\cite{FFLO} or pair-density wave\cite{PDW}. Moreover, a Bogolyubov Fermi surface (FS) can be defined in the NH-SC state by the locus of the quasiparticle nodes, that means, at the boundary between the `paired' and `unpaired' regions at $\varepsilon_{{\bf k}}=\Delta_{\bf k}$. The SC FS is different from the nodal line FS occurs in Hermitian superconductors where the SC gap itself vanishes on the normal state FS, i.e., $\Delta({\bf k}_F)=0$.\cite{SCFermisurface} We discuss these aspects in further details in Sec.~\ref{Sec:Discussion}. 


The two eigenvectors of  Eq.~\ref{BCSHamiltonian} are
\bea
|\psi_{{\bf k}+}\rangle =\left( \begin{array}{ c}
 \alpha_{\bf k}   \\
\beta_{\bf k} \\
\end{array} \right), \ \ 
|\psi_{{\bf k}-}\rangle =\left( \begin{array}{ c}
-\beta_{\bf k}  \\
 \alpha_{\bf k}  \\
\end{array} \right), 
\label{eigenPTH}
\eea
where $\alpha_{{\bf k}}=\sqrt{\frac{1}{2}\left(1+\frac{\varepsilon_{\bf k}}{E_{\bf k}}\right)}$, and $\beta_{{\bf k}}=\sqrt{\frac{1}{2}\left(1-\frac{\varepsilon_{\bf k}}{E_{\bf k}}\right)}$. They follow the usual normalization condition $\langle \psi_{\pm}|\psi_{\pm}\rangle=1$, which is related to the constraint $|\alpha_{\bf k}|^2 + |\beta_{\bf k}|^2=1$. 

However, in the NH-SC case the situation changes. Here, in the `paired region' $|E_{\bf k}|\le|\varepsilon_{\bf k}|$ which makes $\beta_{\bf k}$ imaginary. So, it makes more sense to write $\beta_{\bf k}$ as $\beta_{\bf k}=i\sqrt{\frac{1}{2} (\frac{\varepsilon_{\bf k}}{E_{\bf k}}-1)}$. This gives the constrains that $|\alpha_{\bf k}|^2-|\beta_{\bf k}|^2=1$, and $\alpha^2_{\bf k}+\beta_{\bf k}^2=1$. This leads to an essential problem that the eigenstates are not anymore normalized, because $\langle \psi_{\pm}|\psi_{\pm}\rangle= |\alpha_{\bf k}|^2+|\beta_{\bf k}|^2\ne 1$. According to the \PT-symmetric quantum theory, this problem is solved by taking the so-called $\mathcal{CPT}$-inner product of the eigenstates, as defined in the next section.

\subsection{$\mathcal{CPT}-$inner products}\label{Sec.IIC}

According to the quantum theory of \PT-symmetric Hamiltonian,\cite{ben4,benr,benCopp,PToperator1,PToperator2} although this symmetry guarantees real eigenvalues, positive and finite value of the inner product and unitarity of the states require another symmetry. This symmetry is inherent to all \PT-symmetric Hamiltonian, often termed as conjugating property, and denoted by a $\mathcal{C}$ operator. The nature of the $\mathcal{C}$ symmetry may vary from system to system, and there are multiple ways to define it. The Hamiltonian commutes with both $\mathcal{C}$, and $\mathcal{PT}$, and thus naturally with the combined $\mathcal{CPT}$-operator\cite{benr, benCopp} (for Hermitian Hamiltonian $\CPT=1$). $\tilde{c}_{{\bf k}\sigma}$, $c_{{\bf k}\sigma}$ are the creation and annihilation operators of the non-interaction Hamiltonian $H_0$, which is Hermitian, so $\tilde{c}_{{\bf k}\sigma}\equiv c^{\dag}_{{\bf k}}$ in this case. However, the eigenstates $\psi_{{\bf k}\pm}$ and the Bogoliubov operators of the NH-SC transform under the $\mathcal{CPT}$ operator as follows.

For spinless systems, the time-reversal symmetry is simply $\mathcal{T}=\mathcal{K}$, where $\mathcal{K}$ is the complex conjugation operator. We take the parity operator as $\mathcal{P}=\sigma_z$, where $\sigma_z$ is the third Pauli matrix. Then, the $\mathcal{PT}$-conjugate of the eigenvectors in Eq.~\eqref{eigenPTH} are defined as $\langle\psi_{{\bf k}\pm}|_{\mathcal{PT}}=\left(\mathcal{PT}\psi_{{\bf k}\pm}\right)^{T}=\left(\sigma_z\psi_{{\bf k}\pm}^*\right)^{T}$, as written explicitly by 
\bea
\langle\psi_{{\bf k}+}|_{\mathcal{PT}} = \left( \begin{array}{ cc}
\alpha_{\bf k} &\beta_{\bf k}\end{array} \right),
~\langle\psi_{{\bf k}-}|_{\mathcal{PT}} = 
\left(\begin{array}{ cc}
\beta_{\bf k} & -\alpha_{\bf k}\end{array}\right).
\label{eigenPTH2}
\eea
This leads to the $\mathcal{PT}$-inner product of the eigenvectors to be $\langle\psi_{{\bf k}\pm}|\psi_{{\bf k}\pm}\rangle_{PT}=\pm 1$. In other words, the second eigenvector yields a {\it negative}  norm. This can be rectified by introducing the $\mathcal{C}$-operator as 
\bea
\mathcal{C}&=&|\psi_{{\bf k}+}\rangle \langle \psi_{{\bf k}+}|_{\mathcal{PT}}+ |\psi_{{\bf k}-}\rangle \langle \psi_{{\bf k}-}|_{\mathcal{PT}}, \nonumber \\
&=&(\alpha_{\bf k}^2-\beta_{\bf k}^2)\sigma_z + 2 \alpha_{\bf k}\beta_{\bf k}\sigma_x.
\label{Copp}
\eea 
The key properties of the $\mathcal{C}$ operator are:
\begin{subequations}
\bea
&&\mathcal{C}|\psi_{{\bf k}\pm}\rangle = \pm |\psi_{{\bf k}\pm}\rangle, \quad \mathcal{CPT}|\psi_{{\bf k}\pm}\rangle = |\psi_{{\bf k}\pm}\rangle; \label{Ceigenvalues0} \\
\label{Ceigenvalues}
&& [\mathcal{C},H_{\rm MF}]=0,~~~\left[\mathcal{C},\mathcal{PT}\right ]=0,~~~\mathcal{C}^2 = 1.
\eea
\end{subequations}
Note that the eigenvalues of $\mathcal{C}$ are precisely the signs of the $\mathcal{PT}$ norms of the corresponding eigenstates. Thus the new $\mathcal{CPT}$-inner product becomes always positive, i.e, $\langle\psi_{{\bf k}\pm}|\psi_{{\bf k}\pm}\rangle_{\CPT}=1$. 


The Bogoluibov operators for the two eigenvalues $\pm E_{\bf k}$ are defined as $\gamma _{{\bf k}+}$ and $\tilde{\gamma}_{{\bf k}-}$,  
\bea
\gamma_{{\bf k}+} &=&\alpha_{\bf k}c_{{\bf k}\sigma} - \beta_{\bf k}\tilde{c}_{-{\bf k}\bar{\sigma}},\nonumber\\
\tilde{\gamma}_{{\bf k}-} &=&\alpha _{\bf k}\tilde{c}_{-{\bf k}\bar{\sigma}} +\beta_{\bf k}c_{{\bf k}\sigma}.
\label{bgl}
\eea
Their $\CPT$ conjugates are $\tilde{\gamma}_{{\bf k}\pm}=(\CPT)\gamma_{{\bf k}\pm}(\CPT)^{-1}$. Since $\alpha_{\bf k}$, and $\beta_{\bf k}$ are invariant under $\CPT$, it is easy to show that the Bogoliubov operators anticommute, since fermonic operators $c_{{\bf k}\sigma}$, and $\tilde{c}_{{\bf k}\sigma}$ anticommute :
\bea
\{\tilde{\gamma}_{{\bf k}\pm},\gamma_{{\bf k}\pm} \} &=& \alpha^2_{\bf k}\{\tilde{c}_{{\bf k}\sigma}, c_{{\bf k}'\sigma}\} +  \beta^2_{\bf k}\{c_{-{\bf k},\bar{\sigma}}, \tilde{c}_{-{\bf k}',\bar{\sigma}}\},\nonumber\\
&=& (\alpha^2_{\bf k}+  \beta^2_{\bf k})\delta_{{\bf k},{\bf k'}}=\delta_{{\bf k},{\bf k'}},
\eea
since $\{\tilde{c}_{{\bf k}\sigma},c_{{\bf k}'\sigma'}\}=\delta _{{\bf kk}'}\delta_{\sigma\sigma'}$. Similarly, $\{\tilde{\gamma}_{{\bf k}\pm},\gamma_{{\bf k}\mp} \}=0$ as $\{c_{{\bf k}'\sigma},c_{{\bf k}\sigma}\}=\{\tilde{c}_{-{\bf k}'\bar{\sigma}},\tilde{c}_{{\bf k}\bar{\sigma}}\}=0$.  

The thermal average of the Bogoliubov operators yields $\left\langle \tilde{\gamma }_{{\bf k} + }\gamma _{{\bf k} + }\right\rangle=f(E_{{\bf k}})$, and $\left\langle \tilde{\gamma }_{{\bf k} -}\gamma _{{\bf k}-}\right\rangle=f(-E_{{\bf k}})=1-f(E_{{\bf k}})$, where $f(E_{{\bf k}})$ is the Fermi function.

\subsection{Ground state wavefunction}\label{Sec.IID}
We note that the ground state consists of Cooper pairs for ${\bf k}\in \mathfrak{R_1}$, and unpair electrons for ${\bf k}\in \mathfrak{R_2}$. We first focus on the `paired region' $\mathfrak{R_1}$. The vacuum state is $|\psi_0\rangle$. If the wavefunction of a single Cooper pair at ${\bf k}$ is defined by $|\psi_{1{\bf k}}\rangle$, then the second quantization rule between them arises as
\bea
|\psi_{1{\bf k}}\rangle&=&\tilde{\phi}_{\bf k}|\psi_0\rangle ,\nonumber\\
|\psi_0\rangle&=&\phi_{\bf k}|\psi_{1{\bf k}}\rangle=\phi_{\bf k}\tilde{\phi_{\bf k}}|\psi_0\rangle \ ,
\label{psi1}
\eea
where $\tilde{\phi}_{\bf k}$, and $\phi_{\bf k}$ are the creation and annihilation operators for a single Cooper pair, defined in Eq.~\eqref{phi}. The corresponding $\mathcal{CPT}$ conjugates are $\langle\psi_{1{\bf k}}|_{\CPT}=\langle \psi_{0}|\phi_{\bf k}$, and $\langle\psi_0|_{\CPT}=\langle\psi_{1{\bf k}}|\tilde{\phi_{\bf k}}=\langle\psi_{0}|\phi_{\bf k}\tilde{\phi_{\bf k}}$.
Naturally, $|\psi_0\rangle$, and $|\psi_{1{\bf k}}\rangle$ are orthogonal to each other and posses positive, finite inner products, and hence form a Hilbert space.  This can be seen from the following definitions of the $\mathcal{CPT}$ inner products as
\bea
\langle \psi_0|\psi_0\rangle_\CPT &=& 1,\nonumber\\
\langle \psi_{1{\bf k}}|\psi_{1{\bf k}'}\rangle_\CPT &=& \langle \psi_{0}|\phi_{\bf k}\tilde{\phi}_{{\bf k}'}|\psi_{0}\rangle= \delta_{{\bf k},{\bf k}'}\nonumber\\
\langle \psi_{0}|\psi_{1{\bf k}}\rangle_\CPT &=&\langle \psi_{1{\bf k}}|\psi_{0}\rangle_{\CPT}=0.
\label{orthogonal}
\eea
The probability of the pair creation at ${\bf k}$ is $\beta_{\bf k}^2$, and that of not having a pair is $\alpha_{\bf k}^2=1-\beta^2_{\bf k}$. Therefore the ground state wavefunction of a Cooper pair at ${\bf k}\in\mathfrak{R_1}$ is $|\Psi_{1}({\bf k})\rangle=\alpha _{\bf k}|\psi_0\rangle+ \beta _{\bf k}|\psi_{1{\bf k}}\rangle= (\alpha _{\bf k}+ \beta _{\bf k}  \tilde{\phi }_{\bf k})|\psi_0\rangle$. 

In the `unpaired region', the wavefunction at any ${\bf k}\in\mathfrak{R_2}$ is $|\Psi_{2}({\bf k})\rangle=c_{{\bf k}\sigma}^{\dag}|\psi_0\rangle$. Using Eqs.~\eqref{orthogonal}, we can easily show that the $\CPT$ inner products of both wavefunctions give $\langle\Psi_{\nu}({\bf k})|\Psi_{\nu'}({\bf k}')\rangle_{\CPT}=\delta_{{\bf k},{\bf k}'}\delta_{\nu\nu'}$, where $\nu=1,2$. Therefore, $\Psi_1$ and $\Psi_2$ both belong to the same Hilbert space. The total wavefunction is a product function:
\be
|\Psi_{\rm G}\rangle=\prod_{{\bf k'}\in\mathfrak{R_2},\sigma'}c_{{\bf k'}\sigma'}^{\dag}\prod_{{\bf k}\in\mathfrak{R_1},\sigma} (\alpha_{\bf k}+\beta_{\bf k}c_{{\bf k}\sigma}^{\dag}c_{-{\bf k}\bar{\sigma}}^{\dag})|\psi_{0}\rangle.
\label{wf}
\ee
Therefore, $|\Psi_{\rm G}\rangle$ describes the mean-field wavefunction of the NH pair condensation. A similar wavefunction also arises in the case of FFLO superconductivity,\cite{FFLO} consisting of the product of the `paired' and `unpaired' wavefunctions. Using variational principles, we affirm that this wavefunction describes condensation of NG pairs in Appendix.~\ref{Appen-B}. (The `unpaired region' can also be described by the same $\Psi_1$ function by setting $\alpha=1$, and $\beta=0$.)

\section{Results}\label{Sec:Results}

\subsection{Self-consistent SC gap equation} \label{Sec.R1}
The self-consistent BCS gap equation can be obtained in multiple ways; by minimizing the total energy obtained from the Hamiltonian in Eq.~\eqref{HSC1}, or by simply taking the $\mathcal{CPT}$-inner product of the SC fields defined in Eqs.~\eqref{SCGap}. The total energy ($W_G$) can be obtained by taking the $\mathcal{CPT}$ inner product of the Hamiltonian in Eq.~\eqref{HSC1} with respect with the total ground state in Eq.~\eqref{wf} which yields (see Appendix.~\ref{Appen-B} for details) for ${\bf k}<{\bf k}_F$
\bea 
W_G&=& 2\sum_{{\bf k}\in\mathfrak{R_2}} \varepsilon_{\bf k}+ 2\sum_{{\bf k}\in\mathfrak{R_1}}\varepsilon_{\bf k}|\beta _{\bf k}|^2 \nonumber\\
&&+\sum_{{\bf k}{\bf k}'\in\mathfrak{R_1}} V_{{\bf k}{\bf k'}}\alpha_{\bf k} \beta_{\bf k} \alpha_{\bf k'} \beta_{\bf k'}^*.
\label{WG}
\eea
The first term is the additional energy that arises from the `unpaired region' in the NH-SC, and is zero in the Hermitian case. By minimizing $W_G$ with respect to $\beta^*_{\bf k}$ and $\beta_{\bf k}$, we obtain the condensation of the SC fields as
\bea
\Delta_{\bf k}=-\sum_{{\bf k'}\in\mathfrak{R_1}}V_{\bf kk'}\alpha_{\bf k'}\beta_{\bf k'},~~~\tilde{\Delta}_{\bf k} =-\sum_{{\bf k'}\in\mathfrak{R_1}}V_{\bf k'k}\alpha_{\bf k'}\beta_{\bf k'}^*.\nonumber\\
\label{gapeq}
\eea
We can make few observations here. We notice that in both terms the summation index ${\bf k}'$ switches position in $V_{{\bf k}{\bf k'}}$ which is a key ingredient in obtaining H-SC and NH-SC pairings for symmetric ($V_{{\bf k}{\bf k'}}=V_{{\bf k}'{\bf k}}$) and anti-symmetric ($V_{{\bf k}{\bf k'}}=-V_{{\bf k}'{\bf k}}$) potentials. We also note that although $W_{\rm G}$ contains  both paired and unpaired regions, but the only surviving term in the gap function is the paired region. Therefore, the same equation works for both H-SC and NH-SC cases with $\mathfrak{R_1}$ is extended to the entire BZ in the latter case. Eq.~\eqref{gapeq} can be verified by taking the $\mathcal{CPT}$ inner products of the SC fields with the full ground state wavefunction, i.e., $\langle \Psi_{\rm G}|{\phi}_{{\bf k}}|\Psi_{\rm G}\rangle _{\mathcal{CPT}}$, and $\langle \Psi_{\rm G}|\tilde{\phi}_{{\bf k}}|\Psi_{\rm G}\rangle _{\mathcal{CPT}}$, and substituting them in Eqs.~\eqref{SCGap}. This proves that the generalized anti-symmetric pairing interaction leads to the NH pairing instability. Finally, substituting for $\alpha_{\bf k}$, and $\beta_{\bf k}$, and also introducing the temperature dependence, we obtain 
\bea
\vspace{-.1in}
\Delta_{{\bf k}}&=&-\sum_{{\bf k}'\in\mathfrak{R_1}} V_{{\bf k}{\bf k}'} \frac{\Delta _{{\bf k}'}}{2 E_{{\bf k}'}} \tanh{\left(\frac{ E_{{\bf k}'}}{2 T}\right)}.
\label{sc1}
\vspace{-.14in}
\eea
Similar equation is obtained for $\tilde{\Delta}_{\bf k}$, by substituting $V_{{\bf k}{\bf k}'}\rightarrow V_{{\bf k}'{\bf k}}$. For the robustness of the numerical results, we take various forms of  $V_{\bf kk'}$, such as $V_{\bf kk'}=-V_0$, or $V_{\bf kk'}=V_0 \sin{k_x}\cos{k_y}$, as well as  $V_{{\bf k}{\bf k}'}=V_0 g_{{\bf k}} \tilde{g}_{{\bf k}'}$ ($V_0$ is constant) with $g_{\bf k}=i\sin{k_x}\sin{k_y}$. In all cases, we obtain characteristically the same conclusion a discussed below. 

\begin{figure}[t]
\centering
\hspace{-.17in} 
\includegraphics[width=.50\columnwidth]{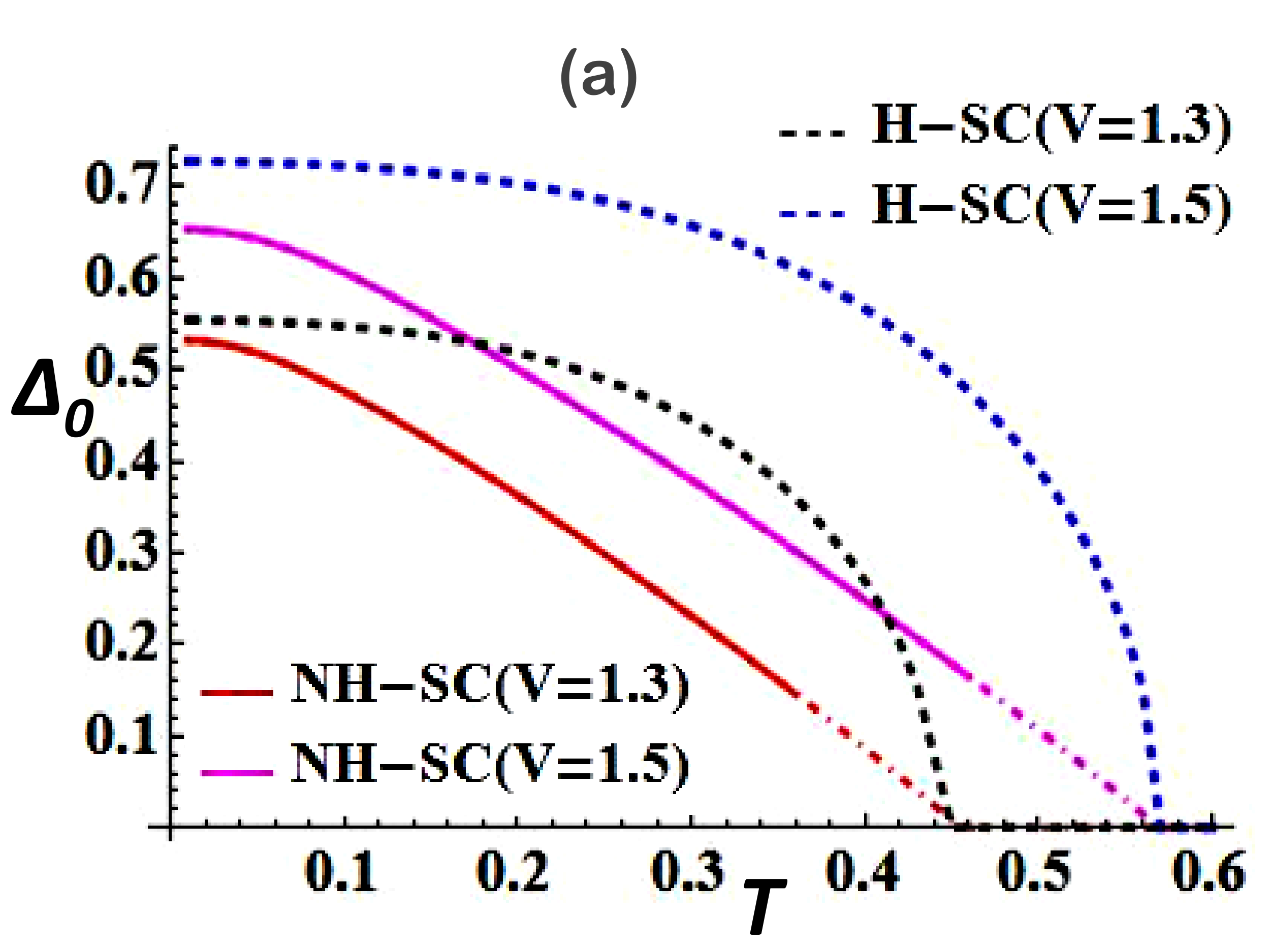}  
\includegraphics[width=.50\columnwidth]{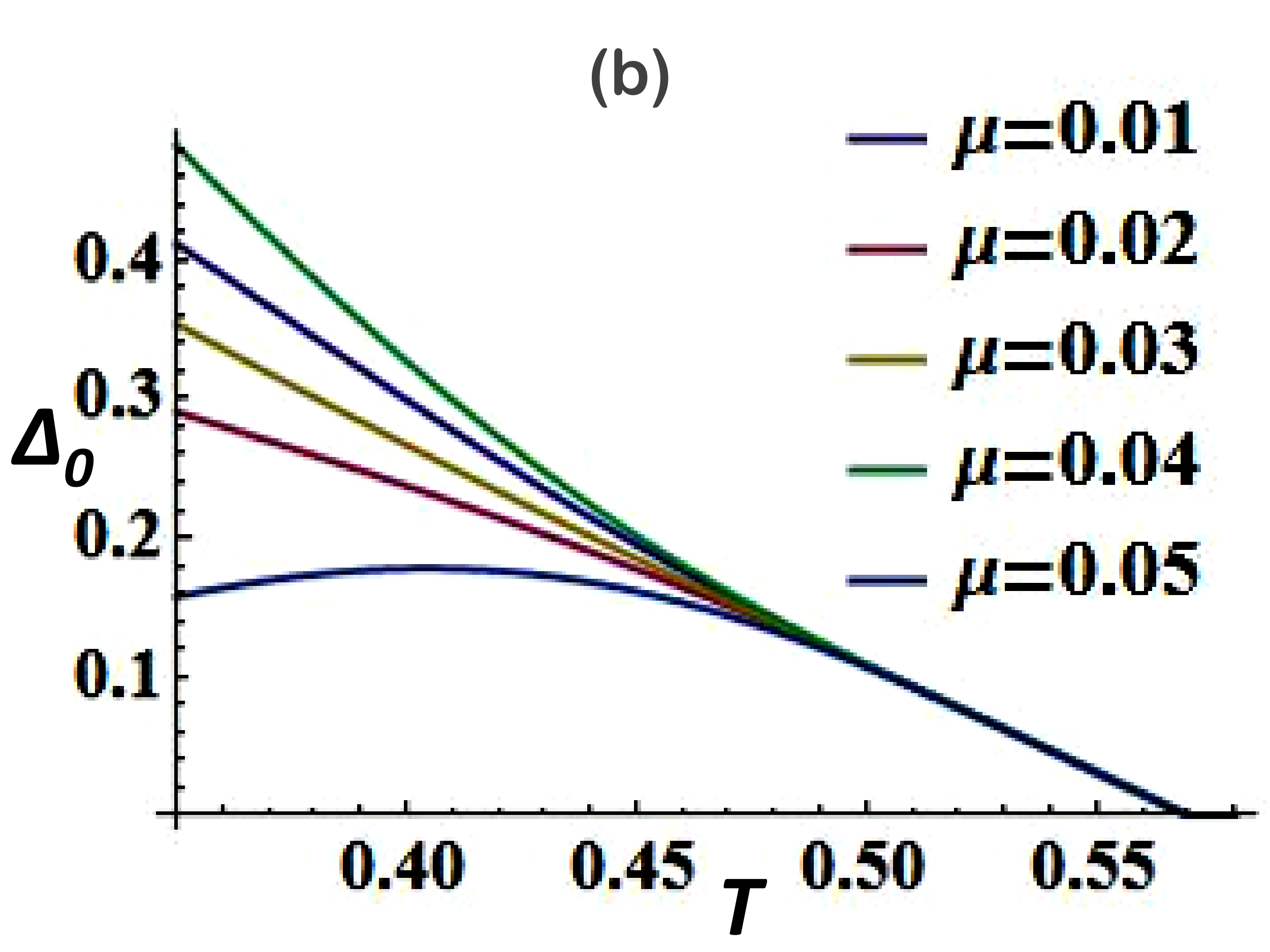} 
\caption{(a-b) Self-consistent values of the SC gap $\Delta_0$ for the NH-SC and H-SC cases, respectively are plotted for different values of the pairing potential, $V_0$, and chemical potential, $\mu$. The NH-SC gap values are plotted in dashed line near $T_c$ to emphasize the fact that due to the first order-phase transition here, the gap discontinuously drop to zero without tracing the smooth curve to reach zero. The exact value where the phase transition occurs is not attainable from the gap function in Eq.~\eqref{sc1}. The temperature where solid to dashed line transition occurs is chosen here arbitrarily and for illustration purpose only.}
\label{fig2}
\end{figure}

Interestingly, the solution of the self-consistent gap equation (Eq.~\eqref{sc1}) gives characteristically different results for the H-SC and NH-SC, keeping all other parameters the same.  In Fig.~\ref{fig2}, we plot self-consistently evaluated SC gap amplitude $\Delta_0 (T)$ for different values of $V_0$ and $\mu$. The H-SC gap shows a typical BCS like temperature dependence with critical exponent 1/2, characterizing a continuous, second order phase transition. In contrast, the NH-SC gap exhibits a linear-in-$T$ dependence near the transition for all values of $V_0$ and $\mu$. We establish below that such a behavior leads to a first-order phase transition, in which the gap discontinuously vanishes, instead of smoothly tracing the dashed line in Fig.~2.

\subsection{Characterization of the phase transition} \label{Sec.R2}

\begin{figure}[t]
\centering
\hspace{-.17in} \includegraphics[width=.52\columnwidth]{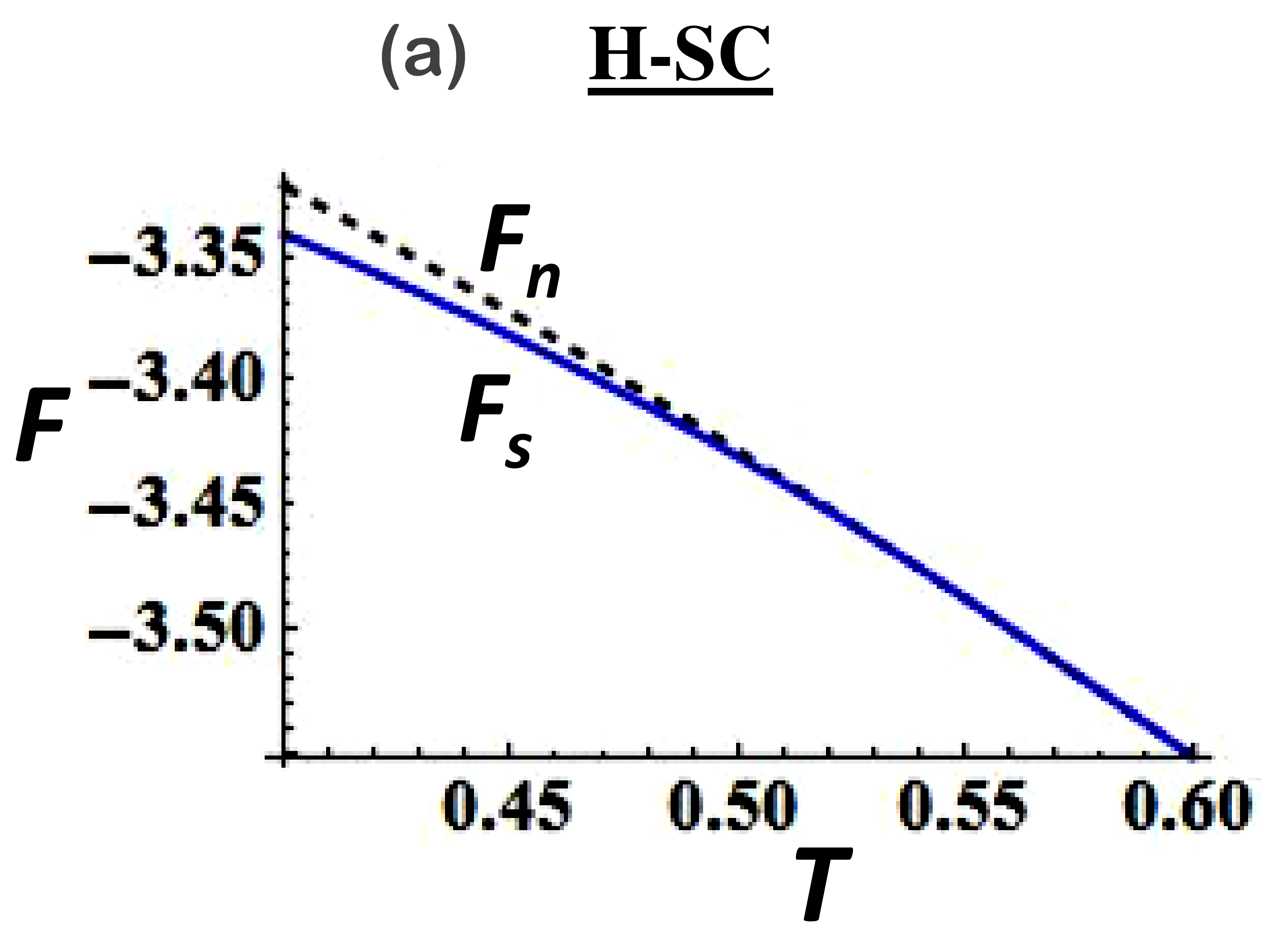}  \hspace{-.065in} \includegraphics[width=.52\columnwidth]{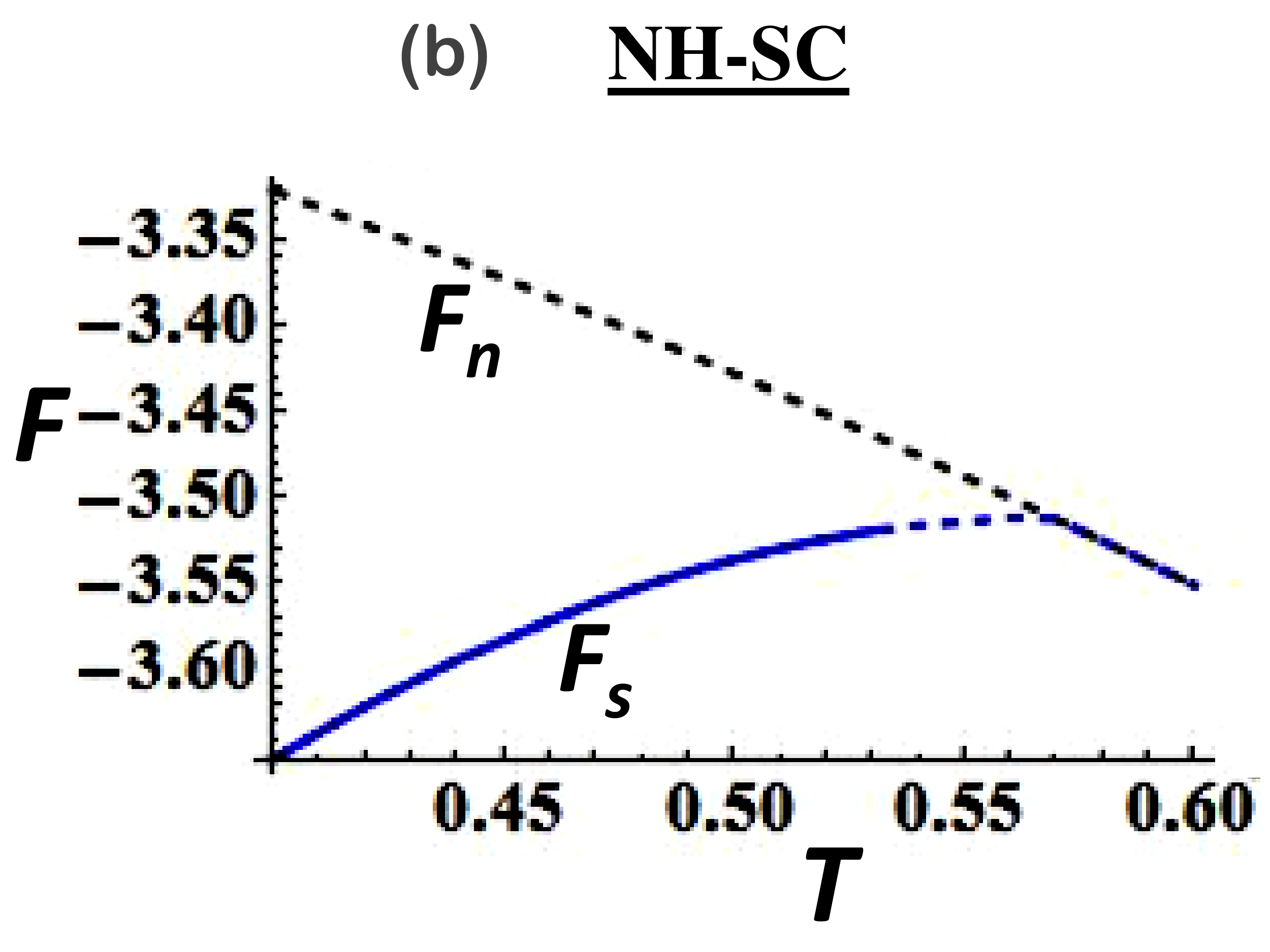} \\
\hspace{-.10in} \includegraphics[width=.49\columnwidth]{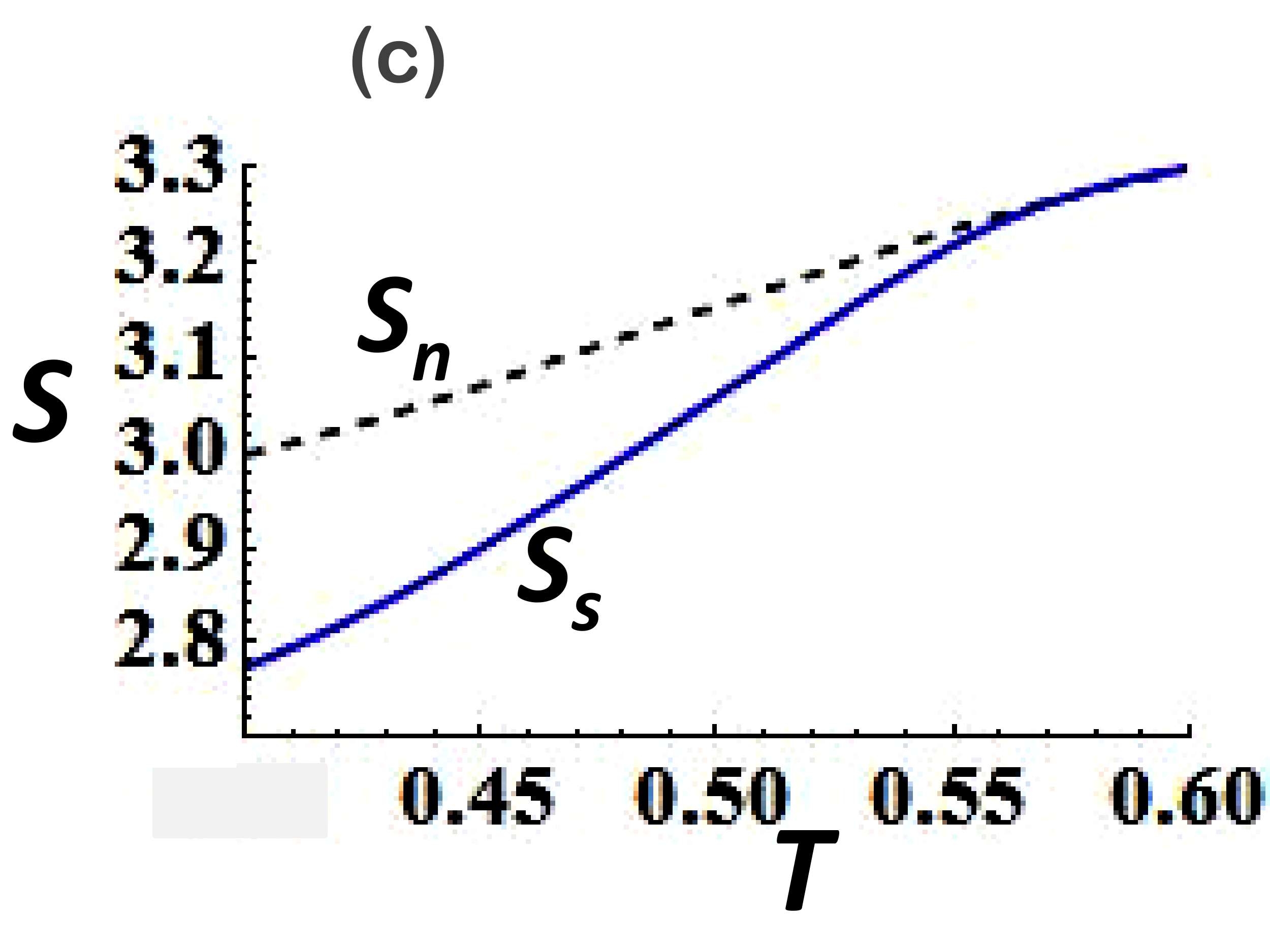}  \hspace{.08in} \includegraphics[width=.49\columnwidth]{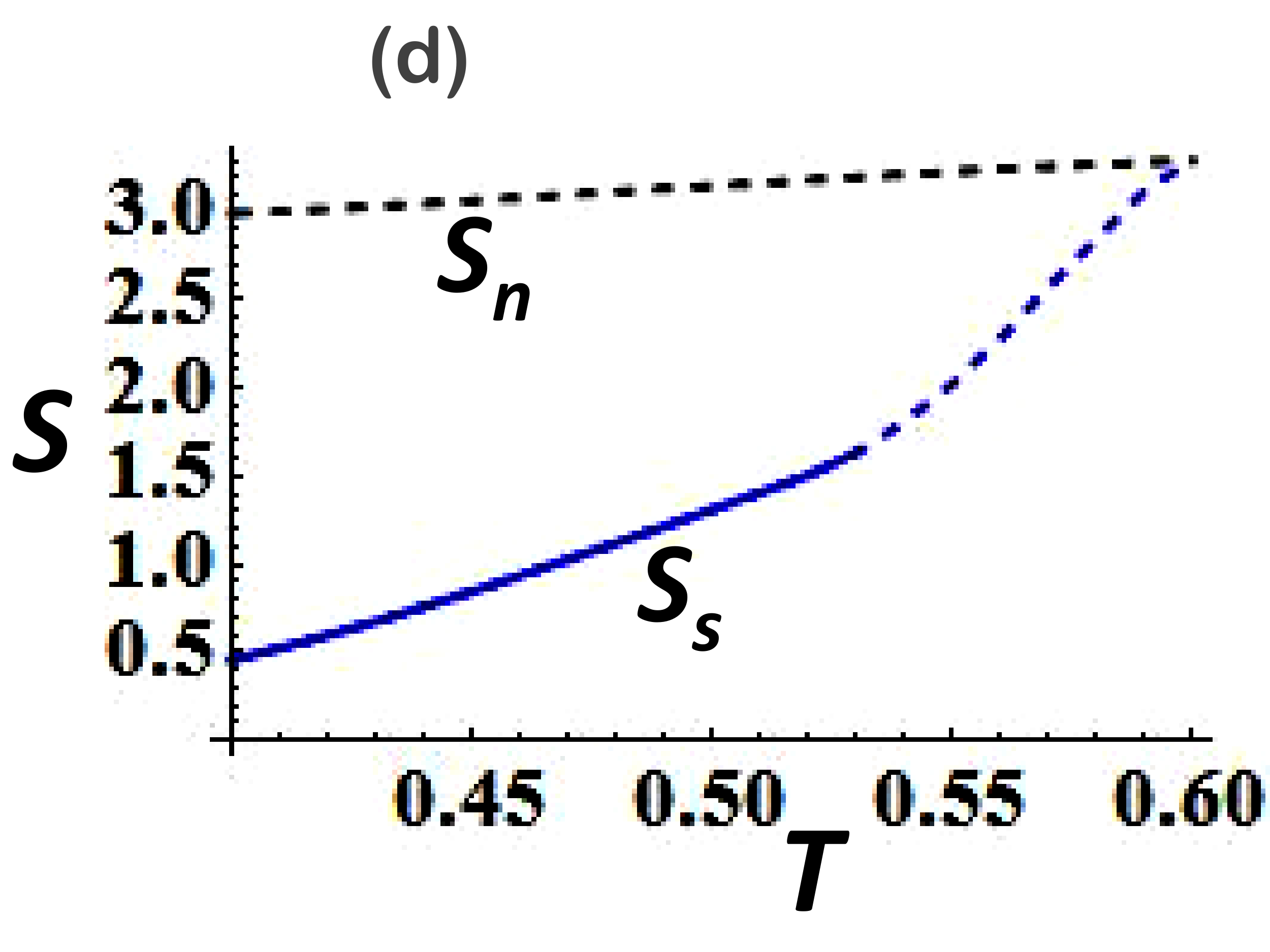} 
\caption{Computed free energy for the H-SC case in (a), and the NH-SC case in (b) are plotted in both normal (dashed line) and SC (solid line) states. Here we choose the same parameter $V=1.5$ and $\mu=.03$ in both cases (in units of $t$). The entropy computed from the  free energies for the H-SC, and the NH-SC cases are plotted in (c) and (d), respectively. Normal and SC free energies obtain similar slope in the H-SC case, and thus its first derivative (entropy) does not obtain any jump, and its second derivative becomes discontinuous at $T_c$ (not shown). On the other hand, in the NH-SC case, these two free energies acquire opposite slope, and thus obtain a discontinuity in the first derivative at $T_c$. The exact location where the jump occurs is arbitrarily chosen, as in Fig. 2. In fact, as mentioned in Fig.~2, the actual phase transition occurs before reaching the predicted $T_c$ from Eq.~\eqref{sc1}. }
\label{fig3}
\end{figure}

Next we delineate the underpinnings of the phase transition by studying the temperature evolution of the free energy and entropy. Implementing the $T$-dependence of the SC gap from Eq.~\eqref{sc1}, we obtain the mean-field Free energy in the SC state [see Appendix~\ref {Appen-B}],
\bea
&&F_s=2\sum_{{\bf k}\in\mathfrak{R_1}} |\varepsilon_{\bf k}|\left [f(E_{\bf k})+(1-2 f(E_{\bf k}))\left(1-\frac{\varepsilon _{\bf k}}{E_{\bf k}}\right)\right]\nonumber\\
&&-\sum_{{\bf k}\in\mathfrak{R_1}}\left(\frac{\Delta_{{\bf k}}\tilde{\Delta}_{{\bf k}}}{2 E_{\bf k}}\right)(1-2f(E_{\bf k}))-TS+2\sum_{{\bf k}\in\mathfrak{R_2}}\varepsilon_{\bf k}f(\varepsilon_{\bf k}),\nonumber\\
\label{fgl}
\eea
where the entropy ($S$) is given as 
\bea
S&=&4\sum_{{\bf k}\in\mathfrak{R_1}}\left[{\rm ln}(1+e^{- E_{\bf k}/T})+ \frac{E_{\bf k}}{T} f(E_{\bf k})\right] \nonumber\\
&&+ 2\sum_{{\bf k}\in\mathfrak{R_2}}\left[{\rm ln}(1+e^{- \varepsilon_{\bf k}/T})\right].
\eea
The last terms in $F_s$ and $S$ give the corresponding contributions from the `unpaired region' in the NH case. We calculate the free energy, and entropy for both H-SC and NH-SC cases, and the results are shown in Fig.~3. For Hermitian and anti-Hermitian superconductors, we obtain $\Delta_{{\bf k}}\tilde{\Delta}_{{\bf k}}=\pm|\Delta_{\bf k}|^2$, which makes the differences in the temperature dependence of $F_s$. As a consequence, we observe that in the NH-SC state, $F_s(T)$ has a opposite slope, compared to that of the normal state free energy $F_n = F_s (\Delta=0)$.  This leads to a `kink' behavior of $F_s$ at $T_c$ which causes a discontinuous jump in the first derivative of the Free-energy, i.e. in the entropy, as shown in Fig.~3(d).  According to the Ehrenfest classification scheme, this phase transition is a first-order type. In the corresponding H-SC counterpart, we recover the second-order phase transition characteristics.   

\subsection {Ginzburg-Landau (GL) framework} \label{Sec.R3}
\begin{figure}[t]
\centering
\hspace{-.16in} \includegraphics[width=0.82\columnwidth]{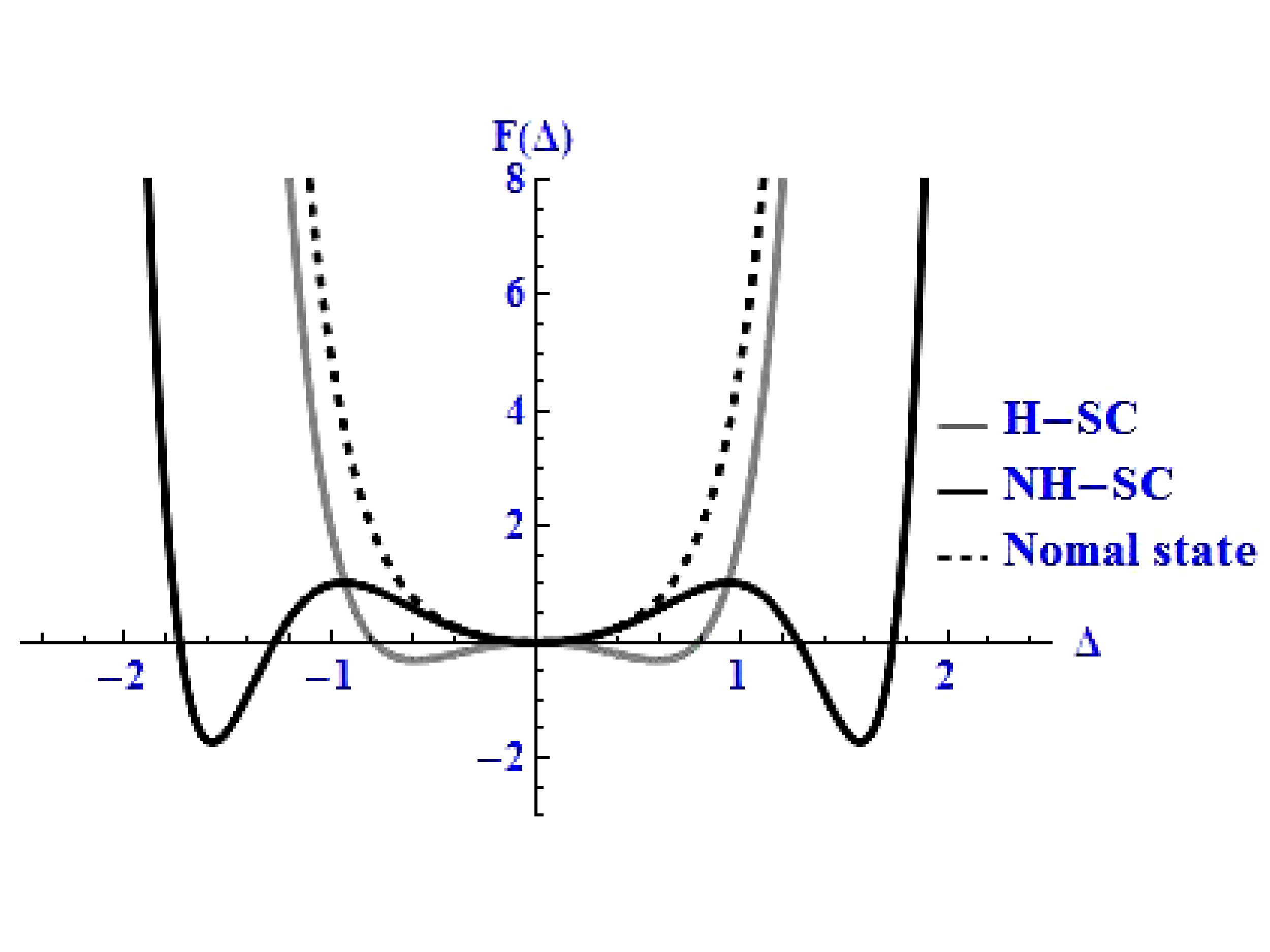} 
\vspace{-.24in}
\caption{Plot of Ginzburg-Landau free energy as a function of the order parameter $\Delta$ for the H-SC and NH-SC cases, with the same parameter sets of $a=-1.5$, $b=1$, $c=2$, and $d=0.5$ (Eq.~\eqref{fgl}). Note that both the first and second orders phase transitions can be reproduced for some values of these parameters. For the H-SC case (gray line), it shows two minima at finite $\pm\Delta_0^c$ and they merge to a single minimum at $\Delta_0=0$ as $a$ smoothly increases and crosses zero. For the same parameter set, the NH-SC system (in which all the odd powers of $|\Delta|^2$ become negative, gives a first order phase transition (black line). This situation is analogous to making $a=-a$, and $c=-c$, while keeping $b$, and $d$ the same. }
\label{GL}
\end{figure}

The conversion of the NH-SC phase transition into a first-order type is parameter-free, and is a direct consequence of the symmetry of the Hamiltonian. Both second, and first order phase transitions in H-SC and NH-SC cases can be reproduced with the same parameter set if we include up to fourth order gap expansion of the Free-energy within the GL theory:
\be
F=F_s-F_n=a(\Delta\tilde{\Delta})+ b(\Delta\tilde{\Delta})^2+c(\Delta\tilde{\Delta})^3+d(\Delta\tilde{\Delta})^4,
\label{fgl}
\ee
where $a$, $b$, $c$ and $d $ are the expansion coefficients. In the H-SC phase, $\Delta \tilde{\Delta}=|\Delta|^2$, so a second order phase transition occurs as $a<0$ below $T_c$, while $b$, $c$, and $d$ are positive. This is evident from the corresponding free energy plot in Fig.~4  in which the free energy minima continuously shift to a finite gap value as $a$ smoothly changes sign.   

Within the lowest order approximation, all these coefficients depend on the normal state properties ($t$, $\mu$, and $V_0$)\ and thus remain very much the same as we switch between H-SC  and NH-SC order parameters [see Appendix.~\ref{Appen-C}]. However, an important change arises from the gap term itself, in that all the odd power of $\Delta\tilde{\Delta}$ terms change sign: $\Delta \tilde{\Delta}=-|\Delta|^2$, and $(\Delta\tilde{\Delta})^3=-|\Delta|^6$, and other terms remain the same. Therefore, for the same set of expansion coefficients, i.e., $a<0$, and $b$, $c$, $d$ are positive, the free energy minima at finite gap value is disconnected to the minimum at $\Delta=0$ through a maximum in between. This situation is equivalent to the parameter values of $a>0$, $b>0$, and $c<0$, and $d>0$ with Hermitian pairing which gives a first order phase transition ($d$ is required to keep the energy bounded). Note that, this first-order transition is slightly different from the one usually obtained in the Hermitian case with $b<0$ with other parameters being positive. 

\subsection{Meissner effect} \label{Sec.R4}
Since magnetic field breaks the $\mathcal{T}$-symmetry, one may expect that there will not be any Meissner effect in the $\mathcal{PT}$-symmetric NH-SC system. However, our calculation shows that in the limit of small $B$, a Meissner effect arises. The $\mathcal{PT}$-symmetric NH Hamiltonian is known to follow a modified continuity equation \cite{bb}. We follow the same strategy for the calculation of current operator with an applied magnetic field \cite{bb}: ${\bf J}({\bf r})=\frac{1}{2}\left[\tilde{\psi}({\bf v}^{\prime}\psi)- ({\bf v}^{\prime}\tilde{\psi})\psi\right ]$, where ${\bf v}^{\prime}={\bf v}-\frac{e{\bf A}}{mc}$, with ${\bf A}$ is the vector potential, and $e$, $m$, $c$ have the usual meanings. The total current can thus be split into paramagnetic, and diamagnetic terms as ${\bf J}={\bf J}_{\rm p} + {\bf J}_{\rm d}$. Fourier transforming the current operators in the corresponding ${\bf k}$, and photon momentum ${\bf q}$-spaces, we get 
\begin{subequations}
\bea
{\bf J}_{\rm p}({\bf q})&=& e\sum_{{\bf k},\sigma}^{\prime}{\bf v}_{k} \tilde{c}_{{\bf k}-{\bf q},\sigma} c_{{\bf k},\sigma},  \label{jp} \\
 {\bf J}_{\rm d}({\bf q})  \label{jd}&=& -\frac{e^2}{ c}{\bf a}({\bf q})\sum^{\prime}_{{\bf k},\sigma} \frac{1}{m^*_{\bf k}} \tilde{c}_{{\bf k}-{\bf q},\sigma} c_{{\bf k},\sigma}.  
\eea
\end{subequations}
Here ${\bf a}({\bf q})$ is the Fourier components of the vector potential, ${\bf v}_{\bf k}$ and $m^*_{\bf k}$ are the band velocity and mass, respectively. 

Both the dispersion $\varepsilon_{\bf k}$ and the SC gaps $\Delta_{k}$, $\tilde{\Delta}_{k}$ involving ${\bf k}$ dependence acquire corrections as the vector potential is turned on. However, as shown in details in Appendix~\ref{Appen-D}, the corrections in $\Delta_{\bf k}$ gives a quadratic-in-${\bf a}$ term in the current term, which we neglect in the present linear-response theory. Therefore, the electromagnetic interaction term is obtained from the kinetic energy only, which yields $H_{\rm int} =-\frac{e}{c}\sum_{{\bf k},{\bf q},\sigma}^{\prime} {\bf v}_{\bf k}\cdot{\bf a}({\bf q})\tilde{c}_{{\bf k}+{\bf q},\sigma} c_{{\bf k},\sigma}+O(a^2)$. With an explicit calculation, we obtain the paramagnetic and the diamagnetic components of the current tensor as:  
\bea
J^{\mu\nu}_{\rm p}(q)&=& -\frac{e^2\beta}{2c^2}a^{\nu}(q)\sum_{{\bf k}\in\mathfrak{R_1}} v_{\bf k}^{\mu}v_{\bf k}^{\nu} {\rm sech}^2\left(\frac{ E_{\bf k}}{2 T}\right)+J_{p,0}^{\mu\nu}, \label{jpmn}\nonumber\\
{J}^{\mu\nu}_{\rm d}(q)&=&\frac{4e^2}{c^2}a^{\nu}(q)\sum_{{\bf k}\in\mathfrak{R_1}} \frac{1}{m^{\mu\nu}_{\bf k}}  \left(1-\frac{\varepsilon _{\bf k} }{E_{\bf k}}\tanh \frac{ E_{\bf k}}{2 T}\right)+J_{d,0}^{\mu\nu},\nonumber\\
\label{jdmn}
\eea
where $\mu$, $\nu$ = $x,~y$. $v_{{\bf k}}^{\mu}=\frac{1}{\hbar}\frac{\partial \varepsilon_{{\bf k}}}{\partial {\bf k}_{\mu}}$, and $\frac{1}{m_{{\bf k}}^{\mu\nu}}=\frac{1}{\hbar^2}\frac{\partial^2 \varepsilon_{{\bf k}}}{\partial {\bf k}_{\mu}\partial {\bf k}_{\nu}}$. $J_{p,0}$ and $J_{d,0}$ are the contributions from the `unpaired regions' which can be obtained by setting $\Delta\rightarrow 0$ in the corresponding currents in the `paired regions'. In the absence of superconductivity here, we can show that $J^{\mu\nu}_{p,0}=-J^{\mu\nu}_{d,0}$, implying that there is no Meissner effect in the unpaired region.

In Fig.~5, we plot the corresponding response kernels in the limit of $q\rightarrow 0$ as $K_i(q) =-\frac{4 \pi J_i(q)}{a(q)}$ ($i$ = p, d). The total response kernel is related to the superfluid density ($n_s$) as $K(0)=\lambda^{-2}=4\pi n_se^2/mc^2$, where $\lambda$ is the penetration depth. We note some characteristic differences between the H-SC and NH-SC cases. For the former case, the diamagnetic term is very much temperature independent, while in the NH-SC counterpart, it reduces smoothly across $T_c$. This is because the momentum values are restricted to the `paired region' in the latter case. Comparing Eqs.~\eqref{sc1} and \eqref{jpmn}, it can be easily deduced that the paramagnetic term is proportional to the SC gap amplitude $\Delta_0$, and thus acquires similar temperature dependence. As a result, near $T_c$, the superfluid density exhibits linear-in-$T$ dependence. As $T\rightarrow 0$, both systems show a linear temperature dependence which is a consequence of the nodal gap structure as seen in other nodal superconductors \cite{DasNodalGap}.

\begin{figure}[t]
\centering
\hspace{-.15in} \includegraphics[width=.5\columnwidth]{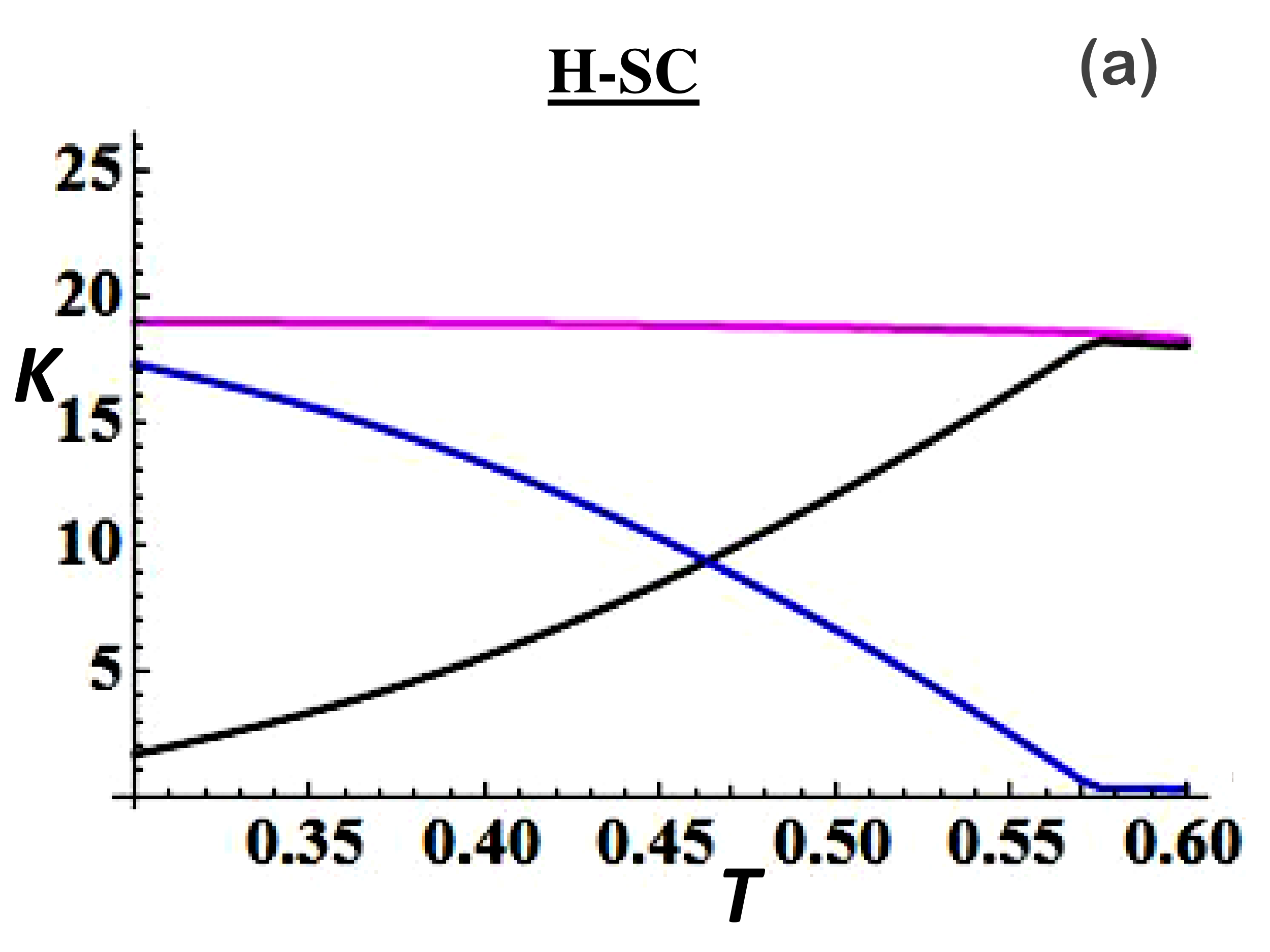}   \includegraphics[width=.5\columnwidth]{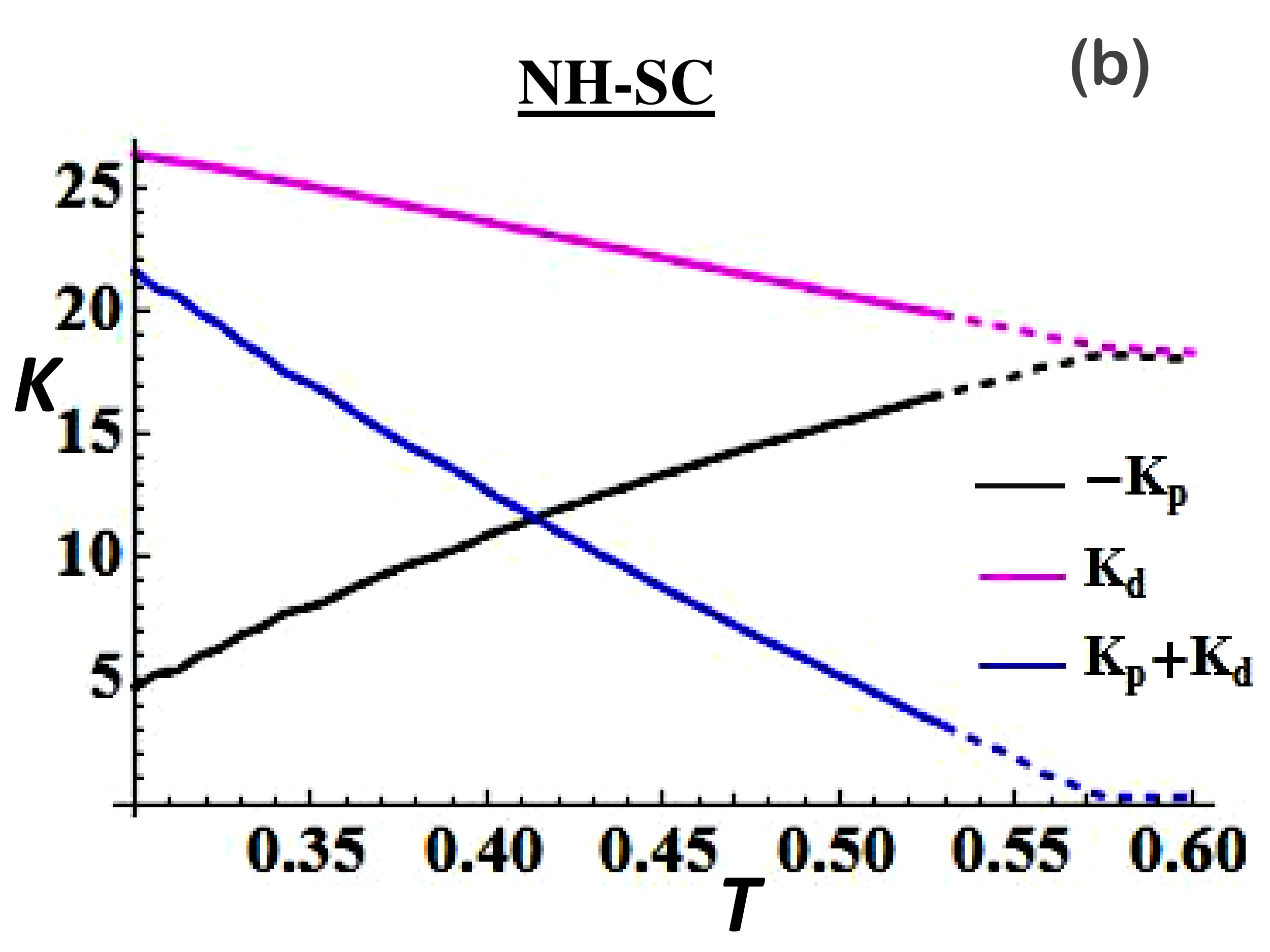} \\
\vspace{-.14in}
\caption{(Color online) Plots of superfluid `Kernels' (proportional to currents) are shown with paramagnetic (black dashed), diamagnetic (magenta), and total (blue) components (along the $xx$ and $yy$-directions), see Eqs.~\eqref{jpmn}, \eqref{jdmn}. The parameter set is kept the same as in Fig.~3. For the NH-SC, all plots are drawn in dashed lines near $T_c$ to emphasize that the location of the discontinuous phase transition is unknown.}
\label{fig4}
\end{figure}

\section {Discussions \& conclusions}\label{Sec:Discussion}

Here we discuss in further details how the NH-SC pairing emerges from a Hermitian normal state, and makes a proper physically realizable quantum phase of matter. We also discuss their potential observations. 

(i) We obtained the general property that the dual requirements of NH and \PT symmetric superconductivity prescribe an {\it anti-Hermitian} Cooper pairing (Eq.~\eqref{DefineNH}) (this conclusion is drawn for a single band superconductor).  This, however, does not lead to an anti-Hermitian Hamiltonian in Eq.~\eqref{BCSHamiltonian}, but a \PT-symmetric NH Hamiltonian which gives real quasiparticle energy in the \PT-invariant region. The SC state in the \PT-broken, `unpaired' region would decay with a rate $\tau\sim\hbar/2\langle E_{\bf k}\rangle$ for ${\bf k}\in\mathfrak{R}_2$, and becomes a normal state with energy $E_{\bf k}=\varepsilon_{\bf k}$. It should be noted that the BCS Hamiltonian in Eq.~\eqref{BCSHamiltonian} itself commutes with the \PT-operator in both `paired' and `unpaired' regions. But its eigenstates are no-longer the eigenstates of the \PT-operator in the `unpaired' region, due to complex eigenvalues. This is a consequence of the anti-linear property of the \T-operator. 

(ii) Interestingly, the NH-SC pairing occurs away from the normal state FS when $|\varepsilon_{\bf k}|\ge|\Delta_{\bf k}|$, unlike in H-SC where superconductivity occurs at all states including on the FS. Recently, in iron-based Hermitian superconductors, it is demonstrated both experimentally and theoretically that superconductivity forms in `insulating bands' which lie close to the Fermi level, but do not cross it.\cite{HDing,Hirschfeld, FCZhang} This implies that if a NH-SC state forms in the low-lying `insulating bands' with a gap, say $\delta$, there may not arise any unpaired region in the BZ if $\delta>\Delta$ at all momenta.    

Another interesting property of the NH-SC phase is that the boundary between the paired and unpaired region is denoted by zero quasiparticle energy states (see Fig.~1). This means, one can obtain a FS even in the SC state. For a H-SC case, this occurs at the locus of $\Delta_{{\bf k}_F}$ where ${\bf k}_F$ is the Fermi momenta and they are called nodal states. In the NH-SC state, additional nodal states occur when $\xi_{\bf k}=\Delta_{\bf k}$.  Recently, the existence of Bogoliubov FS is predicted in H-SC states with line nodes.\cite{SCFermisurface} As mentioned above, if the NH-SC state occurs in an `insulating' band, and that the `insulating' gap $\delta>\Delta$, it may also escape having a Bogoliubov FS (i.e. no unpaired region would then arise). Another special situation may arise when the NH-SC gap function possess a pairing symmetry such that $\Delta_{{\bf k}_F}=0$ at all $k_F$, then the normal state Fermi surface and the Bogoliubov FS will merge.   

(iii) The NH pairing breaks the gauge symmetry, as a consequence of the anti-linear property of the \PT-operator. Let us assume that the Cooper pair field $\phi_{\bf k}$ is transformed by  a homogeneous phase ($\theta$) as $\phi_{\bf k}\rightarrow \phi_{\bf k}e^{i\theta}$, then its \PT-conjugate component obtains $\tilde{\phi}_{\bf k}\rightarrow \tilde{\phi}_{\bf k}e^{-i\theta}$. The interaction potential (Eq.~\eqref{HSC1}) possessing both $\tilde{\phi}_{\bf k}\phi_{\bf k}$ remains invariant under gauge transformation, while the mean-field Hamiltonian in Eq.~\eqref{BCSHamiltonian} is no longer invariant under this transformation. It is easy to show that the system possess particle-hole symmetry and the energy eigenvalues are $\pm E_{\bf k}$ in both H-SC and NH-SC cases. The particle-hole symmetry of the BdG Hamiltonian  is defined by the operation $\Theta H_{\rm MF}({\bf k})\Theta^{-1}=-H^*_{\rm MF}(-{\bf k})$. This condition is satisfied for an antiunitary particle-hole operator $\Theta=\sigma_x\mathcal{K}$, where $\mathcal{K}$ is the complex conjugation.

(iv) It is known that any observable in a NH system is represented by an operator $\mathcal{A}$ which satisfies $\mathcal{A}^{\textrm T}=(\mathcal{CPT})^{-1}\mathcal{A}(\mathcal{CPT})$, where `T' represents the usual `transpose' operation.\cite{PToperator1,PToperator2} This condition guarantees that the $\mathcal{CPT}$ expectation value $\mathcal{A}$ is real, and is preserved under the time-evolution if the Hamiltonian satisfies $H^{\textrm T}=H$. Our mean-field Hamiltonian in Eq.~\eqref{BCSHamiltonian} satisfies this condition. Therefore, the NH-SC state is a proper physical phase which can be realized in condensed matter systems. 

(v) NH-SC state is an emergent quantum phase which is separated by a first-order phase transition from the normal state. The pairing should be \PT-symmetric. In a non-centrosymmetric system, since the system does not have inversion symmetry, the SC order parameter also looses parity. However, to preserve the \PT-symmetry, the \T-symmetry also has to be broken in the SC state in such a way that the system is invariant under their combination. The \T-symmetry does not necessarily have to be broken in the normal state, but it must be broken at the SC phase transition.

(vi) Dzyloshinskii-Moriya (DM) interaction\cite{DM} is an antisymmetric, Hermitian potential. As shown in Appendix~\ref{DM}, it can drive an anti-Hermitian pairing in multiple conditions. If the momentum-conservation principle is relaxed, and a non-local SC potential is obtained as $V_{\bf kk'}=V({\bf k}-{\bf k'})$, Eq.~\eqref{SCGap} implies that an anti-Hermtian pairing naturally emerges from the real DM interaction. Such a case arise when the system is connected to a `momentum-bath' or boosted. Otherwise, if the the particle-number is not conserved in the system, a case that arises in non-equilibrium or when the system is connected to a `number' bath, the expectation values of the SC field $\langle \tilde{\phi}_{\bf k}\rangle$, $\langle \phi_{\bf k}\rangle$ are no-longer Hermitian conjugate to each other, but can be made \PT-symmetric. This gives \PT-symmetric pairs. Finally, in other cases, where the potential itself is made anti-Hermitian, we can obtain an anti-Hermitian, \PT-symmetric pairing. Such conditions requires additional device setup, such as connecting the system to a bath, or open quantum systems or driven systems, as done in optical lattices\cite{opt1,eqv1} or metamaterials\cite{Alaeian, PTmat} to engineer the \PT symmetric conditions.

Electromagnetic metamaterials\cite{Alaeian,PTmat} and optically pumped systems \cite{opt1,eqv1} are two classes of dynamical systems where the realization of $\mathcal{PT}$ symmetric NH Hamiltonians is widely explored. Interestingly, superconductivity is also recently observed in both of these material classes\cite{SCMetamatSn, SCMetamat,SCMetamatAl, OpticalSC}. In metamaterials, since the dielectric function becomes negative along some of the spatial directions, it can lead to directional dependent attractive Coulomb interaction. This raises the possibility of unconventional superconductivity with momentum dependent order parameter \cite{SCMetamat}. Recently, in hyperbolic metamaterials obtained in the mixture of tin and barium nanoparticles \cite{SCMetamatSn}, and in aluminum thin film grown on Al$_2$O$_3$ \cite{SCMetamatAl}, a characteristic enhancement of the superconducting transition temperature ($T_c$), compared to their bulk values, is observed. Again, in optically excited cuprate materials, a significant enhancement of the coherent superconducting transport up to the room temperature is also reported \cite{OpticalSC}. These results suggest that owing to the external drive, unconventional superconductivity arises with its salient properties which deviate from the typical BCS paradigm. 

Various open quantum systems such as optically driven materials, or proximity induced systems are also potential hosts of the NH-SC pairs, if the system has no inversion symmetry. We found that the Free energy for the NH-SC is lower than that of the H-SC for the same magnitude of the order parameter. Therefore, in a suitable condition for a $\mathcal{PT}$-symmetric pairing (such as with a antisymmetric pairing potential), one can expect a NH pairing is more favorable than the H-counterpart.

\begin{acknowledgments} 
We are thankful to T. V. Ramakrishnan, H. R. Krishnamurthy, S. Sachdev for useful discussions. AG acknowledges the financial support from Science and Engineering Research Board (SERB), Department of Science \& Technology (DST), Govt. of India for the National Post Doctoral Fellowship. TD acknowledges the financial support from the same board under the Start Up Research Grant (Young Scientist).
\end{acknowledgments} 

\noindent
\newline\\

\begin{appendix}

\appendix

\section{DM interaction}\label{DM}
We found in Sec.~\ref{Sec:Model} that an anti-Hermitian superconductivity can arise from an anti-symmetric pairing potential. Here we show how an anti-Hermitian pairing can be obtained from an anti-symmetric potential. One of the known anti-symmetric potential is the DM interaction which arises in non-centrosymmetric materials or at the surface of single crystals.  The most general form of the DM interaction is
\be
H_{\rm I}=i\sum_{\bf q}{\bf V}({\bf q})\cdot\left({\bf S}_{\bf q}\times{\bf S}_{-\bf q}\right),
\label{DM1}
\ee
where ${\bf S}_{\bf q}=\sum_{\bf k}\tilde{c}_{{\bf k}\sigma }\vec{\bf \sigma}_{\sigma,\sigma'}c_{{\bf k}+{\bf q}\sigma'}$, with $\sigma$ being the spin index. Potential ${\bf V}({\bf q})$ arises from charge potential gradient is a real and anti-symmetric function, i.e., ${\bf V}({\bf q})=-{\bf V}(-{\bf q})$. Since $S_y$ contain imaginary `i', Eq.~\eqref{DM1} is a Hermitian Hamiltonian. Without loosing generality, for the present 2D case, we fix the electric field direction along the $z$-axis. Therefore, substituting $S^x_{\bf q}$, and $S^y_{-\bf q}$, in the above equation, we obtain
\begin{widetext}
\bea
H_{\rm I}&=&\sum_{\bf q}V_z({\bf q})\sum_{{\bf k}{\bf k}'}\left[(\tilde{c}_{{\bf k}\uparrow}c_{{\bf k}+{\bf q}\downarrow} +  \tilde{c}_{{\bf k}\downarrow}c_{{\bf k}+{\bf q}\uparrow})
(\tilde{c}_{{\bf k}'\uparrow}c_{{\bf k}'-{\bf q}\downarrow} -  \tilde{c}_{{\bf k}'\downarrow}c_{{\bf k}'-{\bf q}\uparrow}) \right.\nonumber\\
&&~~~~~\qquad\qquad \left. - (\tilde{c}_{{\bf k}\uparrow}c_{{\bf k}+{\bf q}\downarrow} -  \tilde{c}_{{\bf k}\downarrow}c_{{\bf k}+{\bf q}\uparrow})
(\tilde{c}_{{\bf k}'\uparrow}c_{{\bf k}'-{\bf q}\downarrow} +  \tilde{c}_{{\bf k}'\downarrow}c_{{\bf k}'-{\bf q}\uparrow})\right]\nonumber\\
&=&-2\sum_{\bf q}V_z({\bf q})\sum_{{\bf k}{\bf k}'}\left[\tilde{c}_{{\bf k}\uparrow}c_{{\bf k}+{\bf q}\downarrow}\tilde{c}_{{\bf k}'\downarrow}c_{{\bf k}'-{\bf q}\uparrow}
- \tilde{c}_{{\bf k}\downarrow}c_{{\bf k}+{\bf q}\uparrow}\tilde{c}_{{\bf k}'\uparrow}c_{{\bf k}'-{\bf q}\downarrow}\right].
\label{DM2}
\eea
\end{widetext}
We denote $2V_z({\bf k})=V({\bf k})$. We notice that the second term is same as the first term when we substitute ${\bf q}\rightarrow -{\bf q}$, $V(-{\bf q})=-V({\bf q})$, and interchange between ${\bf k}$ and ${\bf k}'$ indices. So, we can ignore the second term and extend the ${\bf q}$ summation over the entire Brillouin zone. We also notice that $H_{\rm I}^{\dagger}=H_{\rm I}$, i.e., it contains its own Hermitian conjugate. It is  convenient to rearrange the above equation according to the SC fields as
\bea
H_{\rm I}&=&\sum_{{\bf k}{\bf k}'{\bf q}}V({\bf q})\tilde{c}_{{\bf k}\uparrow}\tilde{c}_{{\bf k}'\downarrow}c_{{\bf k}+{\bf q}\downarrow}c_{{\bf k}'-{\bf q}\uparrow}.
\label{DM3}
\eea
We consider here zero center-of-mass momentum pairing, i.e., ${\bf k}'=-{\bf k}$. In Eq.~\eqref{phi}, we defined the two SC fields $\tilde{\phi}_{{\bf k}} =\tilde{c}_{{\bf k}\sigma }\tilde{c}_{-{\bf k}\bar{\sigma}}$, and $\phi_{{\bf k}} =c_{-{\bf k}\bar{\sigma}}c_{{\bf k}\sigma}$ for particle-particle and hole-hole pairs. The fields are Hermitian conjugate to each other. In terms of the SC fields, we now get $H_{\rm I}=\sum_{{\bf k}{\bf q}}V({\bf q})\tilde{\phi}_{\bf k}\phi_{-{\bf k}-{\bf q}}$. It will be convenient to substitute ${\bf k}+{\bf q}=-{\bf k}'$ which gives
\bea
H_{\rm I}&=&-\sum_{{\bf k}{\bf k}'}V({\bf k}+{\bf k}')\tilde{\phi}_{\bf k}\phi_{{\bf k}'}.
\label{DM5}
\eea
Eq.~\eqref{DM5} gives two solutions if we obey the momentum conservation or not. The corresponding solutions give Hermitian or anti-Hermitian pairings, respectively. In the first case, we define the SC gaps as
\begin{subequations}
\bea
\tilde{\Delta}_{\bf k} &=& -\sum_{{\bf k}'}V({\bf k+k'})\langle \tilde{\phi}_{{\bf k'}}\rangle,\\
\label{SCGap1a}
\Delta_{\bf k} &=& -\sum_{{\bf k}'}V({\bf k'+k})\langle {\phi}_{{\bf k'}}\rangle.
\label{SCGap1b}
\eea
\end{subequations}
Then both the SC gap and the mean-field Hamiltonian are Hermitian, if $V({\bf k+k'})$ is Hermitian.

{\bf Anti-Hermtian pairing:} Next we assume the formation of a pairing gap at a shifted momentum from the pair-field. Here we substitute $V({\bf q})=V({\bf k}-{\bf k}')$. Then the two non-local SC gaps equations are
\begin{subequations}
\bea
\tilde{\Delta}_{\bf k} &=& -\sum_{{\bf k}'}V({\bf k-k'})\langle \tilde{\phi}_{{\bf k'}}\rangle,\\
\label{SCGap2a}
\Delta_{\bf k} &=& -\sum_{{\bf k}'}V({\bf k'-k})\langle {\phi}_{{\bf k'}}\rangle.
\label{SCGap2b}
\eea
\end{subequations}
In this case, we can see that if $V{\bf k-k'}$ is antisymmetric, one obtains $\tilde{\Delta}_{\bf k}=-\Delta^{\dag}_{\bf k}$ , in other words, the pairing is anti-Hermitian. Such a condition can arise if the system is connected to a `momentum-bath' or boosted.

There are other ways to obtain anti-Hermitian pairing. For example, if the expectation values of the SC fields $\langle \tilde{\phi}_{\bf k}\rangle$, and $\langle \phi_{\bf k}\rangle$ are not Hermitian conjugate to each other, but \PT-symmetric, one obtains an non-Hermitian, \PT-symmetric pair from Eqs.~\eqref{SCGap1a}, \eqref{SCGap1b}. This can occur if the particle-number is not conserved, and/or there is an imbalance between the electron-electron and hole-hole pairs driven by proximity effect or external drive etc. Also if one makes the DM interaction itself anti-Hermitian, Eqs.~\eqref{SCGap1a}, \eqref{SCGap1b} give anti-Hermitian pairings without loosing momentum conservation.

{\bf Hartee and Fock terms} Next, we consider the Hartee and Fock fields. In principle, these two terms can introduce translational symmetry breaking, if present. We here consider the simply homogeneous Hartee and Fock terms which are often calculated within the  Density-Functional Theory calculations. It turns out that the Hartee and Fock terms are characteristically different in the DM interaction, than in a symmetric potential. Since the spin-rotational symmetry is already broken in the DM interaction term, Hartee and Fock terms do not break this symmetry again. In the case of no magnetic moment, they do not break any other symmetry and thus do not lead to a phase transition, rather than adding constant potentials to the total Hamiltonian. 

Since DM interaction vanishes at ${\bf q}=0$, the only possible Hartee term that can arise when ${\bf q}=\pm 2{\bf k}$. Hence we get
\begin{eqnarray}
H_{\rm H} &=&\sum_{\bf kk'q}V({\bf q})\left[\langle \tilde{c}_{{\bf k}\uparrow}c_{{\bf k}+{\bf q}\downarrow}\rangle\tilde{c}_{{\bf k}'\downarrow}c_{{\bf k}'-{\bf q}\uparrow}\delta_{2{\bf k}',{\bf q}}\right. \nonumber\\
&&\qquad\quad~~\left. + \langle\tilde{c}_{{\bf k}'\downarrow}c_{{\bf k}'-{\bf q}\uparrow}\rangle\tilde{c}_{{\bf k}\uparrow}c_{{\bf k}+{\bf q}\downarrow}\delta_{2{\bf k},-{\bf q}}\right]\nonumber\\
%
&=&\sum_{\bf k}\left[\Sigma_{\rm H}({\bf k})\tilde{c}_{{\bf k}\downarrow}c_{-{\bf k}\uparrow}+ \tilde{\Sigma}_{\rm H}({\bf k})\tilde{c}_{{\bf k}\uparrow}c_{-{\bf k}\downarrow}\right].
\label{Hartee}
\end{eqnarray}
Here the Hartee self-energy terms are $\Sigma_{\rm H}({\bf k})= \sum_{\bf k'}V(2{\bf k})\langle \tilde{c}_{{\bf k'}\sigma}c_{{\bf k'}+2{\bf k}\bar{\sigma}}\rangle$, and  $\tilde{\Sigma}_{\rm H}({\bf k})= \sum_{\bf k'}V(-2{\bf k})\langle \tilde{c}_{{\bf k'}\bar{\sigma}}c_{{\bf k'}-2{\bf k}{\sigma}}\rangle$. If there is no net magnetic moment in the system, Hartee energy vanishes.

Finally, the Fock term can be written as
\begin{eqnarray}
H_{\rm F} &=&-\sum_{\bf kk'q}V({\bf q})\left[\langle \tilde{c}_{{\bf k}\uparrow}c_{{\bf k}'-{\bf q}\uparrow}\rangle\tilde{c}_{{\bf k}'\downarrow}c_{{\bf k}+{\bf q}\downarrow} \right. \nonumber\\
&&\qquad\qquad~~\left. + \langle\tilde{c}_{{\bf k}'\downarrow}c_{{\bf k}+{\bf q}\downarrow}\rangle\tilde{c}_{{\bf k}\uparrow}c_{{\bf k}'-{\bf q}\uparrow}\right]\delta_{{\bf k}'-{\bf k},{\bf q}}\nonumber\\
&=&\sum_{\bf k}\left[\Sigma_{\rm F}({\bf k})\tilde{c}_{{\bf k}\downarrow}c_{{\bf k}\downarrow}+ \tilde{\Sigma}_{\rm F}({\bf k})\tilde{c}_{{\bf k}\uparrow}c_{{\bf k}\uparrow}\right].
\label{Hartee}
\end{eqnarray}
Here the Fock self-energy terms are $\Sigma_{\rm F}({\bf k})= -\sum_{\bf k'}V({\bf k}-{\bf k}')\langle \tilde{c}_{{\bf k'}\sigma}c_{{\bf k'}{\sigma}}\rangle$, and $\tilde{\Sigma}_{\rm F}({\bf k})= -\sum_{\bf k'}V({\bf k}'-{\bf k})\langle \tilde{c}_{{\bf k}'\bar{\sigma}}c_{{\bf k'}{\bar{\sigma}}}\rangle$. In the absence of any magnetic ordering, it is easy to see that the Fock term will vanish in the anti-symmetric potential.


\section{Total energy minimization and self-consistent gap equation}\label{Appen-B}
In our considered generalized Hamiltonian $H=H_0+H_{\rm I}$ in Eq.~\eqref {HSC1}, the kinetic part can be written as,
\be
H_0=\sum_{\bf k }\varepsilon _{\bf k} \tilde{c}_{{\bf k}\sigma}c_{{\bf k}\sigma}
\ee
The expectation value of the kinetic energy by using the wavefunction given in Eq.~\eqref {wf} can be obtained as,
\begin{widetext}
\bea
W_{\rm KE}&=&\langle\Psi_{\rm G}|H_0|\Psi_{\rm G}\rangle_{\mathcal{CPT}}\nonumber\\
&=&\sum_{{\bf k} }\prod_{\substack{({\bf k}_1,{\bf k}'_1)\in\mathfrak{R_1}\\ \sigma_1,\sigma'_1}}\prod_{\substack{({\bf k_2},{\bf k_2}')\in\mathfrak{R_2}\\\sigma_2,\sigma'_2}}\varepsilon _{\bf k} \langle \phi_{\rm 0}| (\alpha^*_{{\bf k}_1}+\beta^*_{{\bf k}_1}c_{-{\bf k}_1\bar{\sigma}_1}c_{{\bf k}_1\sigma_1})c_{{\bf k}_2\sigma_2}\Big[\tilde{c}_{{\bf k}\sigma}c_{{\bf k}\sigma}\Big]\tilde{c}_{{\bf k}'_2\sigma'_2}(\alpha_{{\bf k}'_1}+\beta_{{\bf k}_1'}\tilde{c}_{{\bf k}'_1\sigma'_1}\tilde{c}_{-{\bf k}'_1\bar{\sigma'_1}})|\phi_{0}\rangle \nonumber\\
&=&\quad\sum_{{\bf k} }\prod_{\substack{({\bf k}_1,{\bf k}'_1)\in\mathfrak{R_1}\\ \sigma_1,\sigma'_1}}\varepsilon _{\bf k} \langle \phi_{\rm 0}| (\alpha^*_{{\bf k}_1}+\beta^*_{{\bf k}_1}c_{-{\bf k}_1\bar{\sigma}_1}c_{{\bf k}_1\sigma_1})\Big[\tilde{c}_{{\bf k}\sigma}c_{{\bf k}\sigma}\Big](\alpha_{{\bf k}'_1}+\beta_{{\bf k}_1'}\tilde{c}_{{\bf k}'_1\sigma'_1}\tilde{c}_{-{\bf k}'_1\bar{\sigma'_1}})|\phi_{0}\rangle \nonumber\\
&&+\sum_{{\bf k} }\prod_{\substack{({\bf k_2},{\bf k_2}')\in\mathfrak{R_2}\\\sigma_2,\sigma'_2}}\varepsilon _{\bf k} \langle \phi_{\rm 0}|c_{{\bf k}_2\sigma_2}\Big[\tilde{c}_{{\bf k}\sigma}c_{{\bf k}\sigma}\Big]\tilde{c}_{{\bf k}'_2\sigma'_2}|\phi_{0}\rangle \nonumber\\
&=&\sum_{({\bf k}<{\bf k}_F)\in\mathfrak{R_1}}2\varepsilon _{\bf k} \beta_{\bf k}\beta_{\bf k}^*+ \sum_{({\bf k}<{\bf k}_F)\in\mathfrak{R_2}}2\varepsilon_{\bf k}.\\
\eea
\end{widetext}
And, similarly the expectation value of the interaction potential (in Eq.~\eqref {HSC1}) is 
\bea
W_{\rm I}&=&\langle\Psi_{\rm G}|H_{\rm I}|\Psi_{\rm G}\rangle_{\mathcal{CPT}},\nonumber\\
&=&\sum_{\bf k\bf k' }V_{{\bf kk'}} \langle\Psi_{\rm G}|\tilde{\phi }_{{\bf k'}}\phi _{\bf k}|\Psi_{\rm G}\rangle_{\mathcal{CPT}}.\nonumber\\
&=&\sum_{\bf k\bf k'}V_{{\bf kk'}}\langle\Psi_{\rm G}|\phi _{\bf k}|\Psi_{\rm G}\rangle_{\mathcal{CPT}} \langle\Psi_{\rm G}|\tilde{\phi }_{{\bf k'}}|\Psi_{\rm G}\rangle_{\mathcal{CPT}},
\eea
Following the same inner product in both `paired' and `unpaired' regions, we can easily notice that the expectation values of the pair fields only survive in the `paired' region, and gives $\langle\Psi_{\rm G}|\phi _{\bf k}|\Psi_{\rm G}\rangle=\alpha_{\bf k}\beta_{\bf k}$. Therefore, we obtain
\bea
W_{\rm I}&=&\sum_{({\bf k},{\bf k}')\in\mathfrak{R_1}} V_{{\bf k}{\bf k'}}\alpha_{\bf k} \beta_{\bf k} \alpha_{\bf k'}\beta_{\bf k'}^*,
\eea
since $\alpha_{\bf k}$ is real. Thus the total energy is $W_{\rm G}= W_{\rm KE}+W_{\rm I}$ in Eq.~\eqref{WG}. Minimizing $W_{\rm G}$ w. r. to $\beta^*_{\bf k}$, we obtain a self-consistent equation (at zero temperature) as
\be
\frac{\alpha_{\bf k} \beta_{\bf k}}{1-2|\beta_{\bf k}|^2}=-\frac{1}{2\varepsilon_{\bf k}}\sum_{{\bf k'}\in\mathfrak{R_1}} V_{{\bf k}{\bf k'}}\alpha_{\bf k'} \beta_{\bf k'}.
\ee
Substituting $\alpha_{\bf k}$, $\beta_{\bf k}$, and $E_{\bf k}$, we obtain the self-consistent gap equation: 
\bea
\vspace{-.1in}
\Delta_{{\bf k}}&=&-\sum_{{\bf k}'\in\mathfrak{R_1}} V_{{\bf k}{\bf k}'} \frac{\Delta _{{\bf k}'}}{2 E_{{\bf k}'}}, 
\label{sc1_T0}
\vspace{-.14in}
\eea
Similarly, by differentiating the total energy with respect to $\beta_{\bf k}$, we obtain $\tilde{\Delta}_{\bf k}$. This proves that the SC condensation can be well described by the self-consistent gap equation used in the main text.



\section{Derivation of the Ginzburg-Landau expansion coefficients}\label{Appen-C}
The derivation of the Ginzburg-Landau (GL) coefficients is standard and the results turn out to be the same for both Hermitian and non-Hermitian cases. This is because the expansion parameters depend on the normal state parameter ($\xi_{\bf k}$) and the interaction potential $V_{\bf kk'}$ which are same in both cases. The path-integral approach to the derivation of the GL coefficients require a Gaussian integral of the Grassmann variables, which follow anticommutation relation. This is the only place were careful treatment for the $\mathcal{PT}$-symmetric Hamiltonian is required, while the rest of the calculation is standard. We start with a Hamiltonian written in the form of 
\be
H=\sum_{{\bf k}\sigma}\varepsilon_{\bf k} \tilde{c}_{{\bf k}\sigma}c_{{\bf k}\sigma }+\sum_{{\bf k},{\bf k'}}V_{{\bf k}{\bf k'}}\tilde{\phi}_{\bf k'}\phi_{\bf k},
\label{HSC12}
\ee
where the pair creation and annihilation operators are defined in Eqs.~\eqref{phi}. The momentum summation is spanned over the entire BZ, but we will split it into the paired and unpaired region when the SC fields are introduced. Then the action is defined as $S=\int_{0}^{\beta}d\tau \mathcal{L}$, where the Lagrangian density $\mathcal{L}=\partial_{\tau}-H$ (where $\tau$ is the imaginary time at finite temperature). Let us define $c$ and $\tilde{c}$ as  the vectors made of all $c_{{\bf k}\sigma}$ and $\tilde{c}_{{\bf k}\sigma}$ respectively, with ${\bf k}\in\mathfrak{R}_1$-symmetric region. Then the partition function is defined as
\bea
\mathcal{Z}&=&\int \mathcal{D}[c,\tilde{c}] e^{-S[c,\tilde{c}]}.
\eea
We define the SC fields $\Delta_{\bf k}$ ( and $\tilde{\Delta}_{\bf k}$ ) according to Eq.~\eqref{SCGap} in Sec. \ref{Sec:Model}A, 
but without taking the expectation values over $\phi$s. Then we perform the Hubbard Stratonovich transformation to the pair fields $\phi_{\bf k}$, and $\tilde{\phi}_{\bf k}$ as
\be
-V\tilde{\phi}\phi \rightarrow \tilde{\phi}\Delta+ \tilde{\Delta}\phi+\frac{\tilde{\Delta}\Delta}{V}.
\ee
(${\bf k}$-dependence is implied.) The partition function hence takes the form
\be
\mathcal{Z}=\int \mathcal{D}[\Delta,\tilde{\Delta},c,\tilde{c}] e^{-\bar{S}[\Delta,\tilde{\Delta},c,\tilde{c}]}\int \mathcal{D}[\Delta,\tilde{\Delta}] e^{-\int_0^\beta  d\tau \frac{\tilde{\Delta}\Delta}{V}},
\label{Z2}
\ee
where 
\bea
\bar{S}&=&\int_0^\beta d{\tau}\Big[\sum_{{\bf k}\sigma}\Big(\tilde{c}_{{\bf k}\sigma}[\partial_{\tau} -\varepsilon _k]c_{{\bf 
k}\sigma }\nonumber\\
&&~~~~~~~~~~~~~~~ -\tilde{\Delta}_{\bf k}c_{-{\bf k}\bar{\sigma}}c_{{\bf k}\sigma}-\Delta \tilde{c}_{{\bf k}\sigma }c^{\dagger
}_{-{\bf k}\bar{\sigma}}\Big) \Big].
\eea
Here we have inserted back the form of the SC pair fields from Eqs.~\eqref{phi}. By introducing Nambu's spinor for the generalized case, $\psi_{\bf k}=\left(
 \begin{array}{ c c c c }
c_{{\bf k}\sigma }   \\
\tilde{c}_{{-\bf k}\bar{\sigma} }   \\
\end{array} \right)$, $\tilde{S}$ can be expressed as, 
\be
\tilde{S}=\int_0^\beta d\tau \sum_{{\bf k}}\tilde{\psi}_{{\bf k}}\left (\partial_\tau - \bf h_{\bf k}\right )\psi_{{\bf k} },
\ee
with \be
\bf h_k=\left(
 \begin{array}{ c c c c }
\varepsilon _{\bf k} \ \ & \Delta _{\bf k}   \\
\tilde{\Delta} _{\bf k} & -\varepsilon _{\bf k}  \\
\end{array} \right).
\label{GLHamiltonian}
\ee
The first integral in Eq.~\eqref{Z2} is a typical Gaussian integral if $c_{\bf k}$, $\tilde{c}_{\bf k}$ are Grassmann variables, that means they anticommute. This is clearly valid for the Hemitian Hamiltonian. For our non-Hermitian case also, $c_{\bf k}$, $\tilde{c}_{\bf k}$ maintain anti-commutation relation since they represent non-interaction Hermitian fermions. The result is valid even if $\bf h_{\bf k}$ is non-Hermitian. Here we can make a distinction between the paired and unpaired regions. It is clear that the Hamiltonian in Eq.~\eqref{GLHamiltonian} is valid in the paired region, while in the unpaired region its a diagonal Hamiltonian with $\Delta_{\bf k}$. Therefore, we can proceed with the generalized derivation and make this distinction at the end result. The integration over $c$ and $\tilde{c}$ variables yield
\bea
\bar{S}&=& \int_0^{\beta} d\tau {\rm ln} \prod_{\bf k}{\rm det}[\partial_{\tau} - \bf h_{\bf k}],\nonumber\\
&=&\int_0^{\beta} d\tau \sum_{\bf k}{\rm Tr}~{\rm ln}[\partial_{\tau} - \bf h_{\bf k}].
\label{Sbar}
\eea
Now including the second term from Eq.~\eqref{Z2}, we obtain the effective action as
\bea
S_{\rm eff}&=&\int_0^{\beta} d\tau \left[\sum_{\bf k}{\rm Tr}~{\rm ln}[\partial_{\tau} - {\bf h_{{\bf k}}}] - \frac{\tilde{\Delta}\Delta}{V}\right].
\label{Sbar}
\eea
Next we Fourier transform to the Matsubara frequency axis $i\omega_n$ to obtain
\bea
S_{\rm eff}&=&\sum_{{\bf k},n}{\rm Tr}~{\rm ln}[i\omega_n - {\bf h}_{\bf k}] - \int_0^{\beta} d\tau \frac{\tilde{\Delta}\Delta}{V},\nonumber\\
&=&\sum_{{\bf k},n}{\rm Tr}~{\rm ln}[\mathbf{G}^{-1}({\bf k},i\omega_n)] - \int_0^{\beta} d\tau \frac{\tilde{\Delta}\Delta}{V}.
\label{Seff}
\eea
We define the $2\times 2$ BCS Green's function (known as Gorkov-Green's function) $\mathbf{G}^{-1}({\bf k},i\omega_n) = i\omega_n-\bf h_{\bf k}$. $\mathbf{G}$ can also be split into the diagonal, non-interacting Green's function $\mathbf{G}_0$ and the off-diagonal SC gap matrix $\mathbf{\Delta}$ as $\mathbf{G}^{-1}=\mathbf{G}_0^{-1}-\mathbf{\Delta}$, where 
\bea
 \mathbf{G}_0&=&\left(
 \begin{array}{ c c c c }
\frac{1}{iw_n-\varepsilon _{\bf k}} \ \ & 0   \\
0& \frac{1}{iw_n+\varepsilon _{\bf k}}  \\
\end{array} \right), \label{G0} \\
{\bf \Delta}&=&\left(
\begin{array}{ c c c c }
0 & \Delta_{\bf k}  \\
\tilde{\Delta}_{\bf k} & 0\\
\end{array} \right).
\label{delta}
\eea
(The momentum and frequency dependencies are implied). Then from Eq.~\eqref{Seff}, we get 
\bea
S_{\rm eff}&=&\sum_{{\bf k},n}{\rm Tr}~{\rm ln}\left[\mathbf{G}_0^{-1}(1-\mathbf{G}_0\mathbf{\Delta})\right] - \int_0^{\beta} d\tau \frac{\tilde{\Delta}\Delta}{V},\nonumber\\
&=& \sum_{{\bf k},n}{\rm Tr}~{\rm ln}\left[\mathbf{G}_0^{-1}\right] -  \sum_{{\bf k},n,l}\frac{{\rm Tr}[\mathbf{G_0\Delta}]^l}{l!} - \int_0^{\beta} d\tau \frac{\tilde{\Delta}\Delta}{V}, \nonumber \\
\label{Seff2}
\eea
where $l$ is integer. The first term just gives a constant shift and can be neglected. All the odd powers of $\Delta$ vanishes due to symmetry. Note that $\mathbf{G}_0$ is defined for the non-interaction ground state which is a Hermitian system and thus it is spanned over the entire BZ in both H-SC and NH-SC cases. $\Delta$ and $\tilde{\Delta}$ are the average gap values over the paired regions. Therefore, only the surviving terms in the Free energy are
\bea
F&=&S_{\rm eff}/\beta 
\approx a (\Delta\tilde\Delta )+b (\Delta \tilde \Delta )^2+c (\Delta \tilde \Delta )^3+d (\Delta \tilde \Delta )^4+... \nonumber \\
\eea
Given that $\mathbf{G_0}$ and $\mathbf{\Delta}$ are diagonal and off-diagonal terms (see Eqs.~\eqref{G0}, \eqref{delta}), their product can be easily evaluated and the final form of the GL coefficients are
\bea
a(T)&=&-\frac{1}{V}-\frac{1}{\beta}\sum_{{\bf k},n}\mathbf{G}_0^{11}({\bf k},i\omega_n)\mathbf{G}_0^{22}({\bf k},i\omega_n),\nonumber\\
&=&-\frac{1}{V}-\frac{N(0)}{\beta}\sum_{n}\int \frac{d\varepsilon}{\omega_n^2 + \varepsilon ^2} \ , \nonumber\\
&=&-\frac{1}{V} - N(0)\log {\frac{\Lambda}{T}} \ ; \\
b(T)&=&-\frac{1}{\beta}\sum_{{\bf k},n}\left[\mathbf{G}_0^{11}({\bf k},i\omega_n)\mathbf{G}_0^{22}({\bf k},i\omega_n)\right]^2,\nonumber\\
&=&-\frac{N(0)}{\beta}\sum_{n}\int \frac{d\varepsilon}{(\omega_n^2 + \varepsilon ^2)^2} =-\frac{N(0)}{\pi^2 T^2}(0.875)\zeta (3) \nonumber \\ \\
c(T)&=&-\frac{N(0)}{\beta}\sum_{n}\int \frac{d\varepsilon}{(\omega_n^2 + \varepsilon ^2)^4}\ =- \frac{N(0)}{\pi^6 T^6}(0.62)\zeta (7) \ ; \nonumber \\ \\
d(T)&=&-\frac{N(0)}{\beta}\sum_{n}\int \frac{d\varepsilon}{(\omega_n^2 + \varepsilon ^2)^6} = -\frac{N(0)}{\pi^{10} T^{10}}(0.492)\zeta (11). \nonumber \\ 
\eea
Here $N(0)$ is the density of states at the Fermi level and $\Lambda$ is the energy cutoff. We notice that all the parameters depend on $G_0$, and thus the integration extends to the entire BZ in both H-SC and NH-SC cases. This is the reason, we expect them to remain the same in both cases. These integrals can be evaluated exactly for a parabolic band, but for tight-binding bands, one needs to perform numerical calculations. However, our purpose of showing that all these coefficients only depend on the normal state parameters $V$ and $\varepsilon_{\bf k}$ is served and that they remain the same in both Hermitian and non-Hermitian cases.
 
 \vspace{.2in}
{\bf Energy minimization revisited from GL free energy:}
With the GL free energy for a generalized Hamiltonian (Hermitian or non-Hermitian), we can revisit the self-consistent gap equation of the SC gap by minimizing the free energy. From Eq.~\eqref{Seff}, the effective action for a uniform field can be written as, 
\bea
S_{\rm eff}&=&-\int_0^{\beta} d\tau \frac{\tilde{\Delta}\Delta}{V} + \sum_{{\bf k}, n}\mbox{ln} \left[w_n^2 +\varepsilon _{\bf k}^2+\Delta\tilde{\Delta}\right].
\eea
Minimizing $TS_{\rm eff}$ w.r.to the order parameter $\Delta$, i.e. by $\frac{\partial}{\partial \Delta}\left (TS_{\rm eff}\right)=0$, we obtain 
\bea
&&\frac{\partial}{\partial \Delta}\left [\frac{1}{\beta}\sum_{\bf k n} {\rm ln}[w_n^2+\varepsilon  _{\bf k}^2 +\Delta \tilde{\Delta} ]-\frac{\Delta \tilde{\Delta} }{V}\right ] =0, \nonumber \\
&& \frac{1}{V}=\frac{1}{\beta}\sum_{{\bf k}, n} \frac{1}{w_n^2+E_{\bf k}^2} \  , \ \ \mbox{where} \ E_{\bf k}=\sqrt{\varepsilon  _{\bf k}^2 +\Delta \tilde{\Delta}}. \ \ \ \ 
\eea
This is the same self-consistent gap equation for a momentum-averaged pairing potential as obtained in Eq.~\eqref{sc1_T0} before. 


\section{Current calculation for NH SC}\label{Appen-D}
In this section, we provide the details of the Meissner effect calculation and discuss how the substitution of the canonical momentum in the SC gap can be ignored in the calculations of the current. The paramagnetic ($J_p$) and diamagnetic ($J_d$) currents, given in  Eqs. (9) and (10) in the main text, are 
\bea
{\bf J}_{\rm p}({\bf q})&=& e\sum_{{\bf k},\sigma}{\bf v}_{\bf k} \ \tilde{c}_{{\bf k}-{\bf q},\sigma} c_{{\bf k},\sigma}, \label{jp2} \\
{\bf J}_{\rm d}({\bf q})&=& -\frac{e^2}{ c}{\bf a}({\bf q})\sum_{{\bf k},\sigma} \frac{1}{m^*_{\bf k}} \tilde{c}_{{\bf k}-{\bf q},\sigma} c_{{\bf k},\sigma}.  \label{jd2}
\eea
The momentum summation is spanned over the entire Brillouin zone. As discussed in the main text, there is no Meissner effect in the normal state, i.e. as SC gaps are set to be zero, the para-, and diamagnetic terms cancel each other. This implies that we can only focus on the calculation in the paired region, and then the values in the unpaired region can be obtained by setting $\Delta=0$. Since again, the total contribution is zero, we do not need to bother about evaluating them explicitly. Therefore, without loosing generality we can restrict ourselves to the paired region only (the prime over the summation in the following momentum summation indicates that the summation is restricted to the paired region only). We henceforth set the photon momentum ${\bf q}\rightarrow 0$. The ${\bf k}$-summation can be split into $+\bf k$ and $-\bf k$ terms, in which we respectively get, 

\begin{widetext}
\large {\bea 
\tilde{c}_{{\bf k},\sigma }c_{{\bf k}\sigma }& =& 
  \alpha_{\bf k} ^2 \tilde{\gamma} _{{\bf k}+ }\gamma _{{\bf k} + }+   \beta_{\bf k}  ^2 (1-\tilde{\gamma}_{{\bf k}- }\gamma _{{\bf k} -})+ \alpha_{\bf k}\beta_{\bf k} \gamma _{{\bf k} + } \gamma _{{\bf k} -}+ \alpha_{\bf k} \beta_{\bf k} \gamma _{{\bf k}-}\gamma _{{\bf k} + } \ ;
\label{1st} \\
\tilde{c}_{-{\bf k}\sigma }c_{-{\bf k}\sigma }  &=& 
 \beta_{\bf k} ^2 (1-\tilde{\gamma} _{{\bf k}+ }\gamma _{{\bf k} + })+  \beta_{\bf k} ^2 \tilde{\gamma}_{{\bf k}-}\gamma _{{\bf k} -}+ \alpha_{\bf k} \beta_{\bf k} \gamma _{{\bf k} -}\gamma _{{\bf k} + }+\alpha_{\bf k} \beta_{\bf k} \tilde{\gamma} _{{\bf k}+ }\tilde{\gamma }_{{\bf k} -}\ .
 \label{2nd} 
\eea }
\end{widetext}
Here $\alpha_{\bf k}$ and $\beta_{\bf k}$ are the Bogoliubov coherence factors defined in the main text. Note that ${\bf v}_{-{\bf k}}=-{\bf v}_{\bf k}$, and $m^*_{-{\bf k}}=m^*_{{\bf k}}$.  Using it, we obtain the current terms as:
%
\bea
{\bf J}_{\rm p}(0)&=& e\sum_{{\bf k}\in {\rm 1QBZ},\sigma}^{\prime}{\bf v}_{\bf k}\Big[ \tilde{c}_{{\bf k},\sigma }c_{{\bf k}\sigma } - \tilde{c}_{-{\bf k}\sigma }c_{-{\bf k}\sigma } \Big]\nonumber\\
&=&e\sum_{{\bf k}\in {\rm 1QBZ},\sigma}^{\prime}{\bf v}_{\bf k}\Big [\tilde{\gamma} _{{\bf k}+ }\gamma _{{\bf k}+}-\tilde{\gamma }_{{\bf k}-}\gamma_{{\bf k}-}\Big] \nonumber \\
&=&e\sum_{{\bf k}\in {\rm 1QBZ},\sigma}^{\prime} {\bf v}_{\bf k} \big [f^+_{{\bf k} }-f^-_{{\bf k}}\big ] \ , 
\label{1p2}
\eea
and,
\bea
{\bf J}_{\rm d}(0)&=& -\frac{e^2}{ c}{\bf a}(0)\sum^{\prime}_{{\bf k}\in 1QBZ,\sigma} \frac{1}{m^*_{\bf k}} \Big[\tilde{c}_{{\bf k},\sigma} c_{{\bf k},\sigma}+\tilde{c}_{-{\bf k}\sigma }c_{-{\bf k}\sigma } \Big]\nonumber\\
&=&-\frac{e^2}{ c}{\bf a}(0)\sum^{\prime}_{{\bf k}\in {\rm 1QBZ},\sigma} \frac{1}{m^*_{\bf k}}\Big[1-( \alpha_{\bf k}^2-  \beta_{\bf k} ^2) \nonumber\\
&& \ \ \ \ \ \ \ \ \ \ \ \ \ \ \ \ \ \ \ \ \ \ \ \ \ \times\Big (1-\tilde{\gamma} _{{\bf k}+ }\gamma _{{\bf k}+}-\tilde{\gamma }_{{\bf k}- }\gamma_{{\bf k} -}\Big)\Big],\nonumber\\
&=&-\frac{e^2}{ c}{\bf a}(0)\sum^{\prime}_{{\bf k}\in {\rm 1QBZ},\sigma} \frac{1}{m^*_{\bf k}}\left [1-\frac{\varepsilon _{\bf k}}{E_{\bf k}}(1-f^+_{{\bf k} }-f^-_{{\bf k}})\right ].\nonumber\\
\label{jp3}
\eea
%
Here the summation is restricted to the first quadrant of the Brillouin zone (1QBZ). We have substituted $\alpha_{\bf k}^2+\beta_{\bf k}^2=1$, and $\alpha_{\bf k}^2-\beta_{\bf k}^2-=\frac{\varepsilon_{\bf k}}{E_{\bf k}}$. Also, the thermal average of the Bogoliubov quasiparticles are $\left\langle \tilde{\gamma }_{{\bf k} + }\gamma _{{\bf k} + }\right\rangle=f^+_{{\bf k}}$, and $\left\langle \tilde{\gamma }_{{\bf k} -}\gamma _{{\bf k} -}\right\rangle=f^-_{{\bf k}}$, where $f^{\pm}_{{\bf k}}$ are the Fermi distribution functions for the quasiparticles $E^{\pm}_{\bf k}$ in the presence of magnetic field.

Next, we evaluate $E^{\pm}_{\bf k}$. Note that both the non-interacting dispersion $\varepsilon_{\bf k}$ and the gaps $\Delta _{\bf k}$, $\tilde{\Delta }_{\bf k}$ in Eq.~\eqref{HSC1} in the main text depend on the momentum, and thus obtain corrections as the vector potential is turned on. In the low-field limit, we expand these terms up to the first order `${\bf a}$' as 
%
\bea
\label{Taylor_ek}
\varepsilon_{{\bf k}-\frac{e{\bf a}}{\hbar c}} &=&\varepsilon_{\bf k}- \frac{e}{c}{\bf v}_{\bf k}.{\bf a}({\bf q}),\\
\Delta_{{\bf k}-\frac{e{\bf a}}{\hbar c}} &=&\Delta_{\bf k}- \frac{2e}{c}{\bf u}_{\bf k}.{\bf a}({\bf q}),\nonumber\\
\tilde{\Delta}_{{\bf k}-\frac{e{\bf a}}{\hbar c}} &=&\tilde{\Delta}_{\bf k}- \frac{2e}{c}\tilde{{\bf u}}_{\bf k}.{\bf a}({\bf q}).
\label{Taylor_delta}
\eea
where the quasiparticle ${\bf v}_{\bf k}=\frac{\partial \varepsilon _{\bf k}}{\hbar\partial {\bf k}}$, and the SC gap velocity ${\bf u}_{\bf k}=\frac{\partial \Delta_{{\bf k}}}{\hbar\partial {\bf k}}$, and so on. Including the second terms in Eqs.~\eqref{Taylor_ek} and \eqref{Taylor_delta}, we  obtain the interaction Hamiltonian as,
\bea
H_{int}&=&-\frac{e}{c}\sum^{\prime}_{{\bf k},q}{\bf a}({\bf q}).\Big( {\bf v}_{\bf k} \tilde{c}_{{\bf k}+{\bf q}, \sigma }c_{{\bf k},\sigma } \nonumber \\
&&+{\bf u}_{\bf k}\tilde{c}_{{\bf k}+{\bf q}, \sigma}\tilde{c}_{-{\bf k},\bar{\sigma }}
+ \tilde{{\bf u}}_{\bf k}c_{{\bf k}+{\bf q}, \sigma }c_{-{\bf k},\bar{\sigma } } \Big), 
\eea
and ${\bf a}({\bf q})$ is the Fourier transform of vector potential ${\bf A}({\bf r})$ in the momentum space. Thus the energies for this system ($H=H_0+H_{int}$, where $H_0$ is the usual kinetic part of the Hamiltonian) can be written as
\bea
E_{{\bf k}}^{\pm}&=&E_{{\bf k}}\pm \frac{e}{c}{\bf v}_{\bf k}\cdot{\bf a}+ \frac{2e}{c}\Big({\bf u}_{\bf k} \alpha_{\bf k}\tilde{\beta}_{\bf k}+\tilde{{\bf u}}_{\bf k}\tilde{\alpha}_{\bf k} \beta_{\bf k}\Big)\cdot{\bf a},\nonumber\\
%
&=&E_{{\bf k}}\pm \frac{e}{c}{\bf v}_{\bf k}\cdot{\bf a}+\frac{2e}{c}{\bf w}_{\bf k}\cdot{\bf a}.
\label{Ef1f2}
\eea
where gap velocity is defined as ${\bf w}_{\bf k}=\Big({\bf u}_{\bf k}\frac{\Delta_{{\bf k}}}{2E_{\bf k}}+\tilde{\bf u}_{\bf k}\frac{\tilde{\Delta}_{{\bf k}}}{2E_{\bf k}}\Big)$, which is obtained after substituting for $\alpha_{\bf k}\tilde{\beta}_{\bf k}=\frac{\Delta_{{\bf k}}}{2E_{\bf k}}$, and $\tilde{\alpha}_{\bf k}\beta_{\bf k}=\frac{\tilde{\Delta}_{{\bf k}}}{2E_{\bf k}}$. For the case of purely imaginary order parameter in the H-SC, $\Delta _{{\bf k}}=\Delta _{{\bf k}}^\dagger =-\Delta _{{\bf k}}$, so the ${\bf w}_{\bf k}$ dependent term in Eq.~\eqref{Ef1f2}  drops out. For a s-wave SC case where the SC-gap does not depend on the momentum, the second term drops out. For NH-SC, the second term contributes. Now from Eq.~\eqref{Ef1f2} we can see the Fermi distributions as,
\bea
f^{\pm}_{\bf k}=f(E_{{\bf k}}^{\pm})&=&f(E_{{\bf k}})\pm \frac{e}{c} ({\bf v}_{\bf k}.{\bf a})\left (-\frac{\partial f_{\bf k}}{\partial E_{{\bf k}}}\right )  \nonumber \\
&&\quad\qquad +\frac{2e}{c}\left({\bf w}_{\bf k}\cdot{\bf a}\right)\left(-\frac{\partial f_{\bf k}}{\partial E_{{\bf k}}}\right) .
\label{f2}
\eea
Thus, 
\bea
\label{fkp}
f_{{\bf k}}^+-f_{{\bf k}}^-&=&\frac{2e}{c}({\bf v}_{\bf k}.{\bf a})\left (-\frac{\partial f_{\bf k}}{\partial E_{{\bf k}}}\right ), \\
\label{fkd} 
1-f^+_{{\bf k}}-f_{{\bf k}}^-&=& 1-2f(E_{\bf k})-\frac{4e}{c}\left({\bf w}_{\bf k}\cdot{\bf a}\right)\left(-\frac{\partial f_{\bf k}}{\partial E_{{\bf k}}}\right). \nonumber \\
\eea
Now going back to the expression of $J_p$ in Eq.~\eqref{jp3} and from Eq.~\eqref{fkp} it is easily found that the third term in Eq.~\eqref{f2} cancels and the final form of $J_p$ can be written as in Eq.~\eqref{jdmn} (in the main text). Now for $J_d$ in Eq.~\eqref{1p2} with the above Eq.~\eqref{fkd}, 
\bea
{\bf J}_{\rm p}(0)&=&\frac{2e}{c}\sum_{{\bf k}\in 1QBZ,\sigma}^{\prime}({\bf v}_{\bf k}.{\bf a})\left (-\frac{\partial f_{\bf k}}{\partial E_{{\bf k}}}\right ).
\label{Jp}
\eea
This term does not include the gap velocity $F_{\bf k}$. However, it appears in the diamagnetic term as
\bea
{\bf J}_{\rm d}(0)&=& -\frac{4e^2}{ c^2}{\bf a}(0)\sum^{\prime}_{{\bf k}\in 1QBZ,\sigma} \frac{1}{m^*_{\bf k}}\left [1-\frac{\varepsilon _{\bf k}}{E_{\bf k}}(1-2f_{{\bf k}})\right ]\nonumber\\
&&\hspace{-.22in}+\frac{4e^2}{c^2}{\bf a}(0)\sum^{\prime}_{{\bf k}\in 1QBZ,\sigma} \frac{1}{m^*_{\bf k}}\frac{\varepsilon _{\bf k}}{E_{\bf k}}\left({\bf w}_{\bf k}\cdot{\bf a}\right)\left(-\frac{\partial f_{\bf k}}{\partial E_{{\bf k}}}\right). \ \ \
\label{Jd}
\eea
The last term in Eq.~\eqref{Jd} is proportional to ${\bf a}^2$, and thus this can be neglected in the low-field region. Substituting $\left (-\frac{\partial f_{\bf k}}{\partial E_{{\bf k}}}\right )=\frac{1}{2T}{\rm sech}^2\left(\frac{E_{\bf k}}{2T}\right)$, and $1-2f_{\bf k}={\rm tanh}\left(\frac{E_{\bf k}}{2T}\right)$ in Eqs.~\eqref{Jp}, and \eqref{Jd}, we obtain the Eqs.~(jdmn) in the main text.

\section{Some other possible interaction potential for the $\mathcal{PT}$-symmetric pairing} \label{Appen-E}
\begin{figure}[h]
\centering
\hspace{-.16in} \includegraphics[width=0.75\columnwidth]{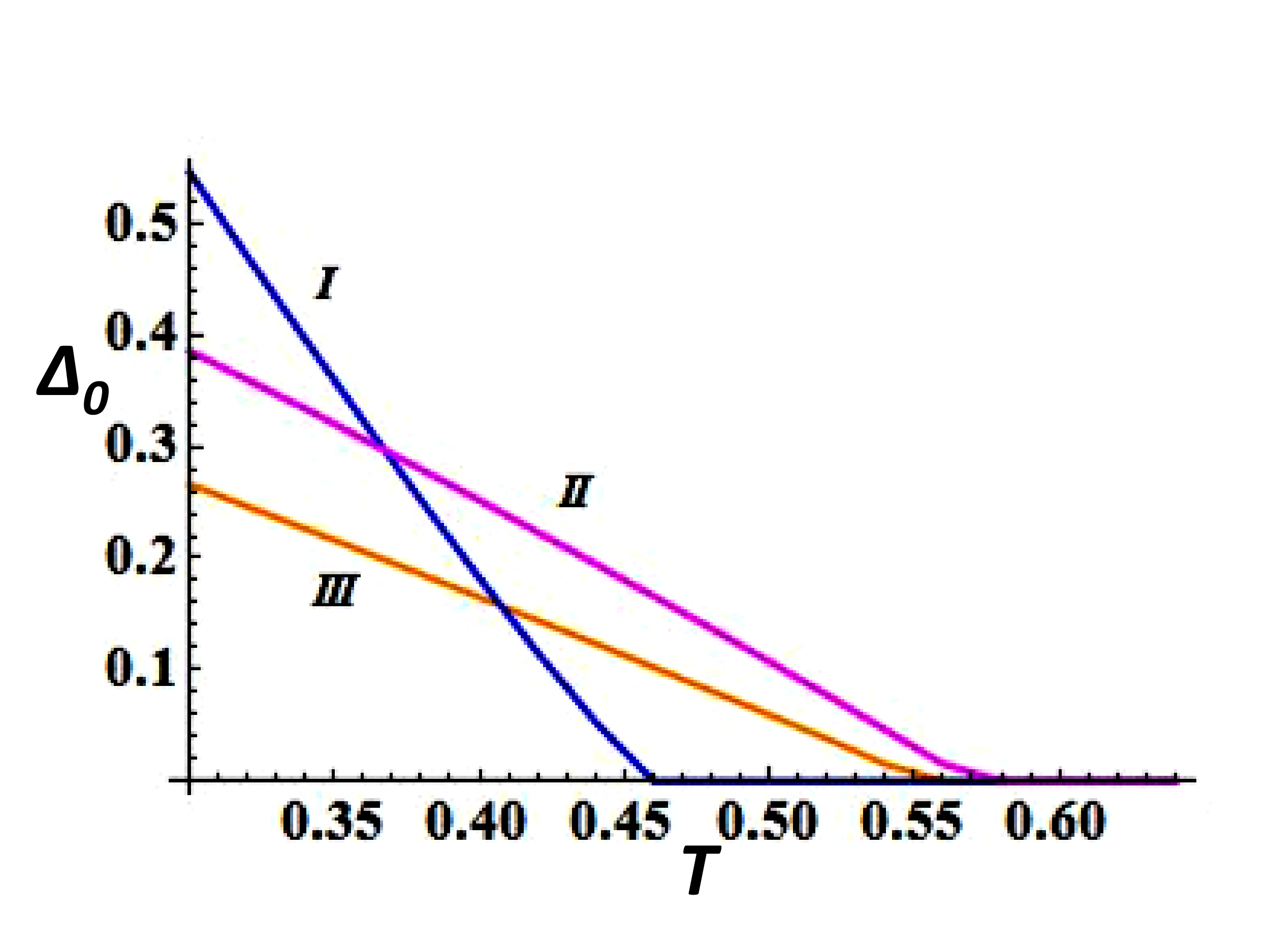} 
\caption{(Color online)  Self-consistent values of the SC gap $\Delta_0$ for different PT-symmetric NH-SC cases are respectively  plotted for the cases (I), (II) and (III), with $V_0=2$ for all cases. For all the cases the temperature dependence of $\Delta_0$ are similar (note that the second case is our considered case in the main text) and thus the other relevant characteristics  of $\mathcal{PT}$-symmetric NH-SC according to our results are also similar.}
\label{3cases}
\end{figure}

We also study some more cases of NH-SC pairing with a pairing potential for which the order parameter is  purely imaginary (breaking $\mathcal{T}$-symmetry), and odd-parity, but invariant under the combined $\mathcal{PT}$-symmetry. We find that the main results and overall conclusions remained unaltered. Different possibilities of pairing potential, $V_{\bf kk'}$ , are considered as: (I) $V_{\bf kk'}=V_0/2$, (II) $V_{\bf kk'}=V_0 \sin {{k}_x} \sin {{k}_y}$, and  (III) $V_{\bf kk'}= V_0 \sin ({{k'}}_x-{k}_x) \cos ({{k'}}_y-{k}_y)$. The gaps for these different cases are compares in the Fig. \ref{3cases}. The nature of the SC gaps show similar behavior for all the NH-SC cases with different pairing potentials.

\end{appendix}


\begin{thebibliography}{1}
\bibitem{dimension}The present theory is developed for two dimensional systems. In one and three dimensions, the form of the order parameter must be altered since the definition of the parity operation is dimension dependent.
\end{thebibliography}
\end{document}